\documentclass{aa}

\usepackage[hyperindex=true,colorlinks=true,citecolor=blue,linkcolor=blue,breaklinks=true]{hyperref}
\usepackage{amsmath}
\usepackage{graphicx}
\usepackage{lscape}
\usepackage{txfonts}
\usepackage{natbib,twoopt}
\usepackage{color}
\usepackage{longtable}

\definecolor{red}{rgb}{1.,0.0,0.}

\definecolor{orange}{rgb}{1.,.65,0.}

\definecolor{vert}{rgb}{.0,.65,0.}

\bibpunct{(}{)}{;}{a}{}{,} 
\newcommandtwoopt{\citeads}[3][][]{\href{http://adsabs.harvard.edu/abs/#3}%
	{\citealp[#1][#2]{#3}}} 
\newcommandtwoopt{\citepads}[3][][]{\href{http://adsabs.harvard.edu/abs/#3}%
	{\citep[#1][#2]{#3}}} 
\newcommandtwoopt{\citetads}[3][][]{\href{http://adsabs.harvard.edu/abs/#3}%
	{\citet[#1][#2]{#3}}}
\newcommandtwoopt{\citeyearads}[3][][]%
{\href{http://adsabs.harvard.edu/abs/#3}{\citeyear[#1][#2]{#3}}}

\newcommand{\dist}{926.3} 
\newcommand{\edist}{5.0} 
\newcommand{\accuracydist}{0.5} 
\newcommand{\parallax}{1.080} 
\newcommand{\eparallax}{0.006} 
\newcommand{\epar}{6} 
\newcommand{\Mcep}{4.859} 
\newcommand{\eMcep}{0.058}
\newcommand{\accuracyMcep}{1.2} 
\newcommand{\Mcomp}{5.141} 
\newcommand{\eMcomp}{0.041}
\newcommand{\accuracyMcomp}{0.8} 

\newcommand{\Mratio}{1.058} 
\newcommand{\eMratio}{0.015}
\newcommand{\Rcep}{31.9} 
\newcommand{\eRcep}{6.0} 
\newcommand{\Mc}{1.546}
\newcommand{\eMc}{0.009}
\newcommand{\Mb}{3.595}
\newcommand{\eMb}{0.033}

\begin{document}

 	\title{Multiplicity of Galactic Cepheids from long-baseline interferometry V. High-accuracy orbital parallax and mass of SU Cygni}
 \titlerunning{Multiplicity of Galactic Cepheids from long-baseline interferometry}
 
 \subtitle{}
 \author{A.~Gallenne\inst{1,2},
 				N.~R.~Evans\inst{3},
 				P.~Kervella\inst{4,5},
 				J.~D.~Monnier\inst{6},
 				C.~R~Proffitt\inst{7},
 				G.~H.~~Schaefer\inst{8},
 				E.~M.~Winston\inst{3},
 				J.~Kuraszkiewicz\inst{3},
 				A.~M\'erand\inst{9},
	 			G.~Pietrzy\'nski\inst{10},
 				W.~Gieren\inst{11},
 				B.~Pilecki\inst{10},
 				S.~Kraus\inst{12},
 				J-B~Le~Bouquin\inst{13},
 				N.~Anugu\inst{8},
 				T.~ten Brummelaar\inst{8},
 				S.~Chhabra\inst{12},
 				I.~Codron\inst{12},
 				C.~L.~Davies\inst{12},
 				J.~Ennis\inst{6}, 
 				T.~Gardner\inst{12},
 				M.~Gutierrez\inst{6},
 				N.~Ibrahim\inst{6},
 				C.~Lanthermann\inst{8},
 				D.~Mortimer\inst{12},
 				\and B.~R.~Setterholm\inst{6}
 }
 
 \authorrunning{A. Gallenne et al.}
 
 \institute{Instituto de Astrof\'isica, Departamento de Ciencias F\'isicas, Facultad de Ciencias Exactas, Universidad Andr\'es Bello, Fern\'andez Concha 700, Las Condes, Santiago, Chile
 	\and  French-Chilean Laboratory for Astronomy, IRL3386, CNRS, Casilla36-D, Santiago, Chile
 	\and Smithsonian Astrophysical Observatory, MS 4, 60 Garden Street, Cambridge, MA 02138, USA
 	\and LESIA, Observatoire de Paris, Universit\'e PSL, CNRS, Sorbonne Universit\'e
 	\and Univ. Paris Diderot, Sorbonne Paris Cit\'e, 5 place Jules Janssen, 92195 Meudon, France
 	\and Astronomy Department, University of Michigan, 941 Dennison Bldg, Ann Arbor, MI 48109-1090, USA
 	\and Space Telescope Science Institute, 3700 San Martin Drive, Baltimore, MD 21218, USA
 	\and The CHARA Array of Georgia State University, Mount Wilson CA 91023, USA
 	\and European Southern Observatory, Karl-Schwarzschild-Str. 2, 85748 Garching, Germany
 	\and Centrum Astronomiczne im. Miko\l{}aja Kopernika, PAN, Bartycka 18, 00-716 Warsaw, Poland
 	\and Universidad de Concepci\'on, Departamento de Astronom\'ia, Casilla 160-C, Concepci\'on, Chile
 	\and Astrophysics Group, Department of Physics \& Astronomy, University of Exeter, Stocker Road, Exeter, EX4 4QL, UK
 	\and Institut de Planetologie et d'Astrophysique de Grenoble, Grenoble 38058, France
 	}

 \offprints{A. Gallenne} \mail{alexandre.gallenne@gmail.com}
 
 
 
 \abstract
 {}
 {We aim at accurately measuring the dynamical mass and distance of Cepheids by combining radial velocity measurements with interferometric observations. Cepheid mass measurements are particularly necessary to help solving the Cepheid mass discrepancy, while independent distance determinations provide crucial test of the period-luminosity relation and Gaia parallaxes.}
 {We used the multi-telescope interferometric combiners Michigan InfraRed Combiner (MIRC) of the Center for High Angular Resolution Astronomy (CHARA) Array to detect and measure the astrometric positions of the high-contrast companion orbiting the Galactic Cepheid SU~Cygni. We also present new radial velocity measurements from ultraviolet spectra taken with the Hubble Space Telescope. The combination of interferometric astrometry with optical and ultraviolet spectroscopy provided the full orbital elements of the system, in addition to component masses and the distance to the Cepheid system.}
 {We measured the mass of the Cepheid, $M_A = \Mcep\pm\eMcep\,M_\odot$, and its two companions, $M_{Ba} = \Mb\pm\eMb\,M_\odot$ and $M_{Bb} = \Mc\pm\eMc\,M_\odot$. This is the most accurate existing measurement of the mass of a Galactic Cepheid (\accuracyMcep\,\%). Comparing with stellar evolution models, we show that the mass predicted by the tracks is higher than the measured mass of the Cepheid, similar to conclusions of our previous work. We also measured the distance to the system to be $\dist\pm\edist$\,pc, i.e. an unprecedented parallax precision of $\epar\,\mu$as (\accuracydist\,\%), being the most precise and accurate distance for a Cepheid. Such precision is similar to what is expected by Gaia for the last data release (DR5 in $\sim 2030$) for single stars fainter than $G = 13$, but is not guaranteed for stars as bright as SU~Cyg.}
 {We demonstrated that evolutionary models remain inadequate in accurately reproducing the measured mass, often predicting higher masses for the expected metallicity, even when factors such as rotation or convective core overshooting are taken into account. Our precise distance measurement allowed us to compare prediction from some period-luminosity relations. We found a disagreement of 0.2-0.5\,mag with relations calibrated from photometry, while relations calibrated from direct distance measurement are in better agreement }
 
 
 \maketitle
 
 %

 \section{Introduction}
 
 One of the most discussed topics in astrophysics today is the tension between the Hubble constant $H_0$ determined for the present universe based on Cepheids and Type Ia supernovae and that from the early universe based on Planck satellite Cosmic Microwave Background observations \citep[see e.g.][]{Freedman_2021_09_3}. Extensive efforts are being made to reduce uncertainties involved, since the discrepancy hints at possible new physics. For Cepheids (the foundation of the local determination), the Leavitt (Period-Luminosity) Law needs to be calibrated to the highest accuracy and we need to understand  the  physics of Cepheid pulsation and evolution as thoroughly as possible. A crucial test of our understanding is the luminosity at the Cepheid stage for a given stellar mass. Evolutionary tracks are beginning to incorporate new insights about rotation in stellar interiors from astroseismology in phases leading up to the Cepheid stage. Thus Cepheid masses of the highest accuracy coupled with very accurate luminosities (partly through Gaia among other approaches) are a stringent test of forefront stellar modelling, and an important underpinning of objects used in deriving $H_0$.
 
It has been known since the first hydrodynamic modelling of Cepheid pulsation in the 1960's that masses inferred from pulsation and evolutionary calculations differed. This difference has now been reduced to $\leq 20$\,\%. The most cited scenarios to explain this discrepancy are a mass-loss mechanism during the Cepheid’s evolution, convective core overshooting, and rotation during the main-sequence stage \citep{Anderson_2014_04_0,Neilson_2011_05_0,Keller_2008_04_0,Bono_2006__0}, but none agrees with the measured dynamical mass \citep{Evans_2024_08_0,Evans_2018_08_1}. This discrepancy points to an uncomfortable lack in our understanding of objects important in the $H_0$ tension. While calibration of Cepheid luminosities will be significantly improved in the final Gaia release, masses can only be directly measured in binary systems, and hence only a limited number are available.
 
 We currently have two capabilities available which have made it possible to directly measure masses for several non-eclipsing Cepheids. High resolution spectra in the ultraviolet (UV) allow the orbital velocity amplitude of hot companions of Cepheids to be measured. Combining this with the orbital amplitude of the Cepheid from the ground-based orbit provides the mass ratio. In addition, interferometry has resolved a number of systems, providing the semi-major axis ($a_1 + a_2$) and the inclination. \object{V1334~Cyg} is a system for which complete observational data are available: an interferometric orbit and a mass ratio from the orbital
 velocity amplitudes for the Cepheid and the companion \citep{Gallenne_2018_11_0}. For this combination both the masses of the stars and a distance can be determined, resulting in the most accurate mass
 and distance to a Galactic Cepheid ($M_A = 4.29 \pm 0.13\,M_\odot, d = 720.4\pm 7.8$\,pc).
 
 This paper is part of a series discussing binary or multiple systems containing Cepheids, and what can be inferred about Cepheid masses from them. \object{SU~Cyg} = HD~186688 = HR~7518 is a $V= 6.44$\,mag Cepheid star with a period of 3.85\,d. It has a asymmetric light curve and a large pulsation amplitude, consistent with pulsation in the fundamental mode. Orbital motion in the system containing the Cepheid SU Cyg was found by \citet{Hellerich_1919_11_0}. An orbit was first determined by \citet{Evans_1988_03_0}. Ground-based velocities of the Cepheid were then combined with velocities of the companion from the International Ultraviolet Explorer (IUE) satellite by \citet{Evans_1990_06_0}. The companion was found to be itself a short-period binary. This produced a dynamical lower limit to the mass of the Cepheid of $5.9\pm 0.4\,M_\odot$.  The stars in the system are thus A (the Cepheid), Ba (the hottest star) and Bb. Inspection of a low resolution International Ultraviolet Explorer (IUE) spectrum showed that the star B has a strong Ga II feature at 1400\AA, indicative of a chemically peculiar HgMn star \citep{Evans_1995_05_0}.  This is consistent with a rotation rate slowed to synchronize with the orbital period of 4.67\,d. This was further discussed by \citet{Wahlgren_1998_04_0}.
 
 In this paper we revisit the SU~Cyg system using new UV spectra from the Hubble space telescope and new interferometric observations with the goal of measuring accurate dynamical masses and distance. We present the observations and the data reduction process in Sect.~\ref{section__observations_and_data_reduction}. In Sect.~\ref{section__data_analysis} we detailed the data analysis to extract radial velocities and astrometric motion, which we then used for a global orbital fit in Sect.~\ref{section__orbital_fitting}. We then discussed our results in Sect.~\ref{section__discussion} and conclude in Sect.~\ref{section__conclusion}.
 
 \section{Observations and data reduction}
 \label{section__observations_and_data_reduction}
 
 \subsection{Ultraviolet spectra}
 
 As with many Cepheids, the hottest star in the SU~Cyg system is the companion. The hottest star outshines the Cepheid in the ultraviolet. Because of this the velocity of the companion can be measured on Hubble space telescope (HST) Space Telescope Imaging Spectrograph (STIS) spectra.
 
 We obtained ultraviolet spectra with the STIS instrument on board the HST in Cycles 22 and 23 (PI: Gallenne) and 28 and 29 (PI: Evans). The high-resolution echelle mode E140H was used in the wavelength region 1163-1357\,\AA. The reduction procedure is described in detail in \citet{Evans_2018_10_0}. The echelle orders are combined into a 1D spectrum using an appropriate blaze function \citep{Baer_2018_01_0}.
 
\begin{figure}[!h]
 	\centering
 	\resizebox{\hsize}{!}{\includegraphics{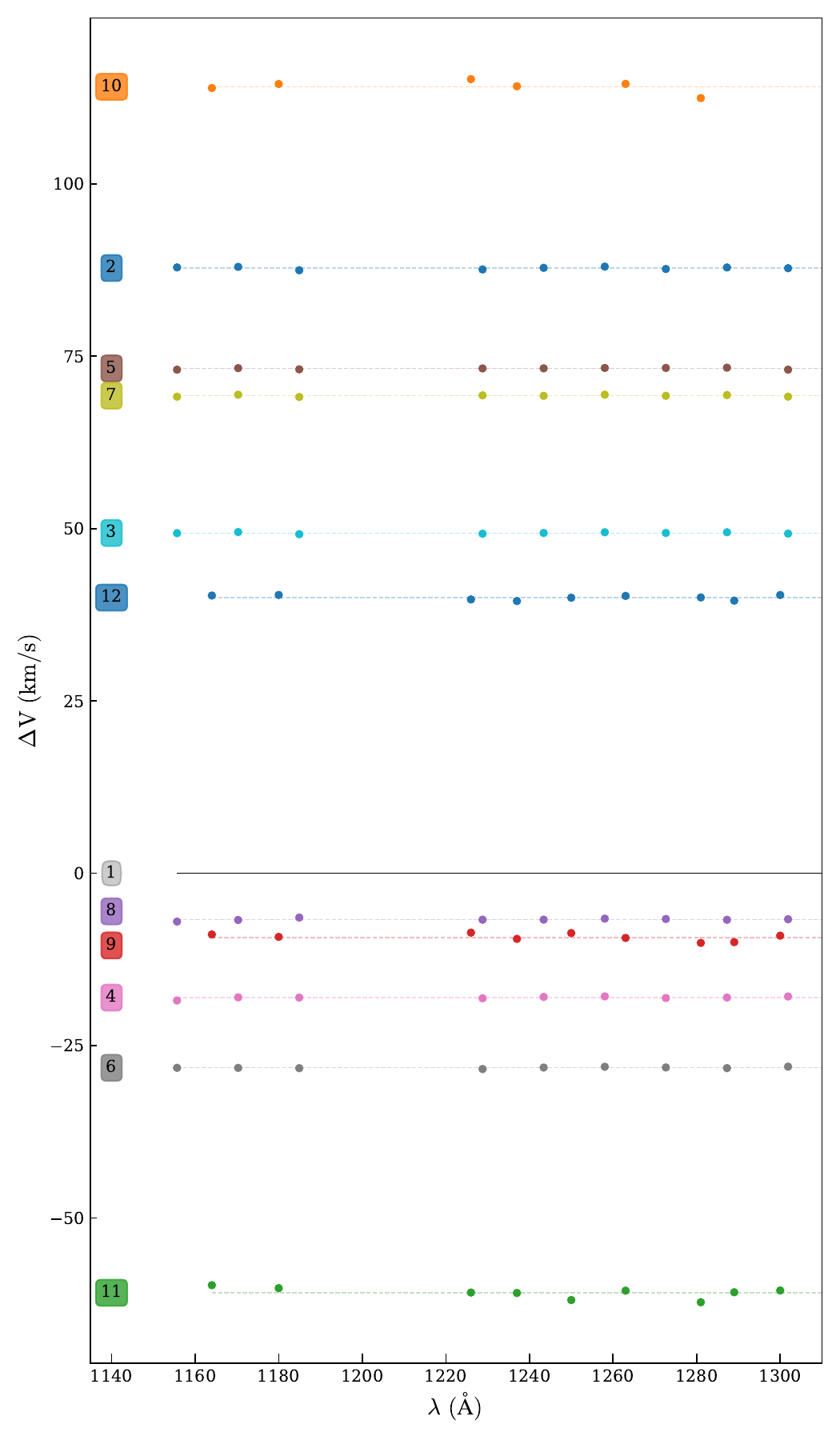}}
 	\caption{Velocities in the wavelength segments for the STIS observations. The observation number from Table~\ref{table_stis_rv} is on the left. Velocities are determined by cross-correlation with the first observation, which is shown as the line at $\Delta V_\mathrm{r} = 0$. The average velocity for each observation is shown as a dashed line.}
 	\label{figure__wavelength_segments}
 \end{figure}

 The velocity differences between the spectra were measured using an IDL cross-correlation routine, as described in  \citet{Evans_2018_10_0}. The first spectrum in the series was arbitrarily selected for the
 velocity difference anchor.  Cross-correlation was  done in wavelength segments typically 10\,\AA\/ wide. Because SU~Cyg B is a sharp-lined HgMn star with many lines, the agreement between segments is very good, as shown in Fig.~\ref{figure__wavelength_segments}.
 
 On the other hand, in spectra with more typical broad lines, such as V1334~Cyg and S~Mus, interstellar lines can be used to check the velocity shift zero point variation due to the positioning of the star within the aperture \citep{Proffitt_2017_09_0}.  In SU~Cyg this cannot be done because the spectrum is full of lines with a similar width to interstellar lines. Internal errors depend largely on the number of lines available and also the rotational velocity of the star. For SU~Cyg~B, internal consistency between segments is excellent as shown by Fig.~\ref{figure__wavelength_segments}.
 
 The observation started with a peak-up on the star. Nominally, an ACQ/Peak is accurate to 5\,\% of the slit width. We used the 0.2X0.09 aperture, and the plate scale of the E140H is 0.047''/pixel in the dispersion direction. A nominal centring accuracy  $0.05 \times 0.09$ / 0.047 is thus about 0.1 pixels. One pixel is $\lambda/228000$, so a tenth of a pixel would give about $0.1c/ 228000$ or about 0.13\,km~s$^{-1}$, which is probably not the limiting accuracy of the wavelength scale. Instead, the accuracy is  probably limited by the stability of the bench and the accuracy of the wavecal measurements. We added extra deep wavecals that bracketed our exposures to help correct for any drifts in that. The CALSTIS pipeline software then interpolates the shift in the velocity alignment as measured from the two lamp exposures to the mean time of the observation. Even if the actual shift in the bench is highly non-linear over the course of an observation, any additional error should be less than half the value of the shift between the two wavecals. We find that the average shift in the position of the lamp spectrum between the leading and trailing wavecal is equivalent to a velocity change about $-0.28\pm0.23$\,km~s$^{-1}$, with the most extreme shift being about $-0.7$\,km~s$^{-1}$. The small scatter in the STIS velocities around the best fit model in the next Section (Fig.~\ref{figure__disentangled_RV}) also confirms the accuracy of the velocities. 
 
 \subsection{Interferometry}
 
 We collected long-baseline optical interferometric data at Georgia State University's Center for High Angular Resolution Astronomy (CHARA) Array \citep{ten-Brummelaar_2005_07_0} located at Mount Wilson Observatory. The CHARA Array consists of six 1m aperture telescopes in an Y-shaped configuration with two telescopes along each arm, oriented to the east (E1, E2), west (W1,W2) and south (S1, S2), offering good coverage of the $(u, v)$ plane. The baselines range from 34\,m to 331\,m, providing an angular resolution down to 0.5\,mas at $\lambda = 1.6\,\mu$m.
 
 Data were collected with the Michigan InfraRed Combiner \citep[MIRC,][]{Monnier_2004_10_0} before 2019 and the upgraded Michigan InfraRed Combiner-eXeter \citep[][]{Anugu_2020_10_0} after 2019. In addition, recent improvements at CHARA include the commissioning of a new 6-telescope beam combiner MYSTIC \citep[Michigan Young STar Imager at CHARA][]{Monnier_2018_07_1,Setterholm_2023_04_0}, designed alongside the MIRC-X upgrade and capable of simultaneous observations. MIRC combined the light coming from all six telescopes in the $H$ band ($\sim 1.6\,\mu$m), with three spectral resolutions ($R = 42, 150$ and 400). The recombination of six telescopes gives simultaneously 15 fringe visibilities and 20 closure phase measurements, that are our primary observables. MIRC-X also combines the light from six telescopes, with the spectral resolution $R = 50, 102$, and 190. Our MIRC and MIRC-X observations used only the lowest spectral resolution to favour the signal-to-noise ratio (S/N). MYSTIC is a $K$-band instrument working similarly to MIRC-X and offering a spectral resolution of 50. The log of our observations is available in Table~\ref{table__log}.

We followed a standard observing procedure, i.e., we monitored the interferometric transfer function by observing a calibrator before and after our Cepheids. The calibrators were selected using the \emph{SearchCal} software\footnote{\url{http://www.jmmc.fr/searchcal}} \citep{Chelli_2016_05_0}  provided by the Jean-Marie Mariotti Center. They are listed in Table~\ref{table__log}.

The data were reduced with the standard MIRC and MIRC-X pipeline\footnote{\url{https://gitlab.chara.gsu.edu/lebouquj/mircx_pipeline.git}} \citep{Monnier_2007_07_0,Anugu_2020_10_0}. The main procedure is to compute squared visibilities and triple products for each baseline and spectral channel, and to correct for photon and readout noises. MYSTIC data are also reduced by the MIRC-X pipeline.

\begin{table*}[!h]
	\centering
	\caption{Journal of the observations.}
	\begin{tabular}{cccccccc}
		\hline
		\hline
		Date  &  HJD\tablefootmark{a}		&	Combiner	& Configuration  & $N_\mathrm{sp}$ & $N_{vis}$  & $N_{CP}$  & Calibrators \\
		\hline
		2016-07-18  & 2457587.814952 & MIRC & S1-E2-W1-W2 & 8 & 156 & 310 &  1  \\
		2016-07-19  & 2457588.783663 & MIRC & S1-S2-E1-E2-W1-W2 & 8 & 426 & 748 & 1,2  \\
		2019-07-14  & 2458678.779086 &  MIRCX & S1-S2-E1-E2-W1-W2  & 8 & 3419  & 5487 & 3,4,5 \\
		2020-06-30  & 2459030.831882 &  MIRCX & S1-S2-E1-E2-W1-W2  &  8 & 398 & 4320 & 5 \\
		2020-07-01  & 2459031.794510 &  MIRCX & S1-S2-E1-E2-W1-W2 & 8 &   1947  & 9600  & 3,4,5 \\
		2021-07-20  & 2459415.767570 &  MIRCX & S1-S2-E1-E2-W1-W2 & 8 & 3738 &  6751   &  3,5,6 \\
		2021-09-19  & 2459476.675433 &  MIRCX & S1-S2-E1-E2-W1-W2 & 8 & 2139 &  2985  & 7  \\
		2022-08-20  & 2459812.476640 &  MIRCX & S1-S2-E1-E2-W1-W2 & 15 & 832 & 833 &  6    \\
		2022-08-20  & 2459812.529639 &  MYSTIC & S1-S2-E1-E2-W1-W2 & 10 & 813 &  527  &  6 \\
		2022-08-21  & 2459812.767333 &  MIRCX & S1-S2-E1-E2-W1-W2 & 15 & 4279 &  6931  &  5,6,8   \\
		2022-08-21  & 2459812.776533 &  MYSTIC & S1-S2-E1-E2-W1-W2 & 10 & 2710 & 4218   & 5,6,8   \\
				\hline
	\end{tabular}
	\label{table__log}
	\tablefoot{$N_\mathrm{sp}$: number of spectral channel. $N_\mathrm{vis}$: number of visibility measurements. $N_\mathrm{CP}$: number of closure phase measurements. The calibrators used have the following angular diameters: 1: $\mathrm{\theta_{LD}(\object{HD189395}) = 0.197\pm0.014}$\,mas, 2: $\mathrm{\theta_{LD}(\object{HD178187}) = 0.244\pm0.017}$\,mas, 3: $\mathrm{\theta_{LD}(\object{HD332518}) = 0.306\pm0.007}$\,mas, 4: $\mathrm{\theta_{LD}(\object{HD227002}) = 0.278\pm0.006}$\,mas, 5: $\mathrm{\theta_{LD}(\object{HD333533}) = 0.303\pm0.007}$\,mas, 6: $\mathrm{\theta_{LD}(\object{BD+28 3437}) = 0.287\pm0.006}$\,mas, 7: $\mathrm{\theta_{LD}(\object{HD332720}) = 0.295\pm0.008}$\,mas, 8: $\mathrm{\theta_{LD}(\object{BD+30 3674}) = 0.272\pm0.006}$\,mas.\\
		\tablefoottext{a}{Converted from JD using the PyAstronomy package \citep{Czesla_2019_06_0}.}
	}
\end{table*}
 
 \section{Data Analysis}
 \label{section__data_analysis}
 
 SU Cyg is part of a triple system which makes its study a bit more difficult. In the following, we define the Cepheid as the primary star A, the hottest star of the companion pair as the secondary B$\mathrm{_a}$ and the tertiary component as B$\mathrm{_b}$. 
 
\subsection{STIS radial velocities}
\label{subsection__stis_radial_velocities}

RVs of the companion were determined using the cross-correlation technique. No synthetic template spectrum was used, instead we estimated a velocity difference with respect to the first observation by cross-correlating each of the spectra against it. This was typically done in 11 segments of approximately 10\AA, which were inspected to optimize the features so that a feature was not divided by a boundary. Mild smoothing was included. These velocity differences are directly used in our orbital model fitting. Measurements are listed in Table~\ref{table_stis_rv}, with internal uncertainties derived from the differences between segments.

\begin{table}[!h]
	\centering
	\caption{Differential Radial Velocities of the component Ba.}
	\begin{tabular}{ccc}
		\hline
		\hline
		\#  &  HJD\tablefootmark{a}	& $\Delta V\mathrm{_B}$	 \\
		&  (day)	&	(km~s$^{-1}$) 	   \\
		\hline
		1  &  2456978.827532  & $  0.00\pm0.04$ \\   
		2  &  2457091.706662  & $ -87.70\pm0.05$ \\
		3  &  2457154.563949  & $ -49.30\pm0.07$ \\
		4  &  2457244.319744  & $  18.10\pm0.06$ \\
		5  &  2457373.543445  & $ -73.20\pm0.07$ \\
		6  &  2457474.382831 & $  28.30\pm0.09$ \\
		7  &  2457584.426308  & $ -69.20\pm0.07$ \\
		8  &  2457665.443512  & $   6.70\pm0.05$ \\
		9  &  2459273.561783  & $   9.30\pm0.16$ \\
		10 &  2459275.481451  & $-113.90\pm0.40$ \\
		11 &  2459656.639142  & $  60.90\pm0.23$ \\
		12 &  2459658.558639  & $ -40.00\pm0.11$ \\
		\hline
	\end{tabular}
	\tablefoot{\tablefoottext{a}{Converted from JD using the PyAstronomy package \citep{Czesla_2019_06_0}.}
	}
	\label{table_stis_rv}
\end{table}

In the ultraviolet wavelength range, the Cepheid is faint compared to its companions, therefore the STIS velocities inform us about the orbital motion of the companions around their common centre of mass (i.e. the system B = B$\mathrm{_a}$+B$\mathrm{_b}$) plus the motion around the centre of mass with the Cepheid. We therefore need to disentangle the centre of mass motion of the short orbital period from the long orbital period.

To disentangle the orbital motions, we used the following model, assuming a circular short-period orbit (consistent hypothesis for such short orbit) and therefore an argument of periapsis $\omega_\mathrm{short} = 0$:
\begin{displaymath}
	V_\mathrm{B} = \gamma + K_\mathrm{B} [\cos(\omega - \pi + \nu) + e \cos(\omega - \pi)] + K_\mathrm{Ba} \cos \nu_\mathrm{Ba}
\end{displaymath}
with $\gamma$ the systemic velocity, $K_\mathrm{B}$ the semi-amplitude of the centre of mass of B in the long orbit, $K_\mathrm{Ba}$ the semi-amplitude in the short orbit, $\omega$ the argument of periastron of B's orbit (with $\pi$ subtracted to take into account the symmetry compared to the equations in Sect.~\ref{section__orbital_fitting}), $\nu$ the true anomaly of the centre of mass in the long orbit, $e$ the eccentricity of the long orbit, and $\nu_\mathrm{Ba}$ the true anomaly in the short orbit.

The true anomaly is defined implicitly with the following Keplerian parameters $P_\mathrm{orb} , T_\mathrm{p} , e$ (respectively for the long- and short-period orbit), and Kepler's equation:
\begin{align*}
	\tan \frac{\nu (t)}{2} &= \sqrt{\frac{1 + e}{1 - e} } \tan \frac{E(t)}{2}\\
	E(t) - e \sin E(t) &= \frac{2\pi (t - T_p)}{P_\mathrm{orb}},
\end{align*}
where $E(t)$ is the eccentric anomaly, $T_p$ the time of periastron passage, $t$ the time of radial velocity observations, and $P_\mathrm{orb}$ the orbital period. In the following, we will use $P_\mathrm{L}, T_\mathrm{p,L}, P_\mathrm{S}$ and $T_\mathrm{p,S}$ to refer to the orbital period and time of periastron passage for the long and short orbit, respectively.

Due to the finite speed of light, we also adjusted at each iteration the times of observations to account for the light travel time effect in the short orbit \citep[LTTE,][]{Irwin_1952_07_0,Irwin_1959_05_0}. The LTTE causes apparent shifts in the timing of the RVs because light from Ba takes time to travel to us as the star moves in its short orbit (LTTE for the long orbit is adjusted later in the global fit). More details can be also found in \citet{Konacki_2010_08_0} or \cite{Zucker_2007_01_0}.

However, the RVs from STIS are differential velocities as they were calculated by cross-correlation with the first spectrum \citep[see][]{Gallenne_2018_11_0}, therefore our model to fit the STIS velocities is:
\begin{displaymath}
	\Delta V_\mathrm{B}(t) = V_\mathrm{B}(t) - V_\mathrm{B}(t_0)
\end{displaymath}

\begin{figure}[!h]
	\centering
	\resizebox{\hsize}{!}{\includegraphics{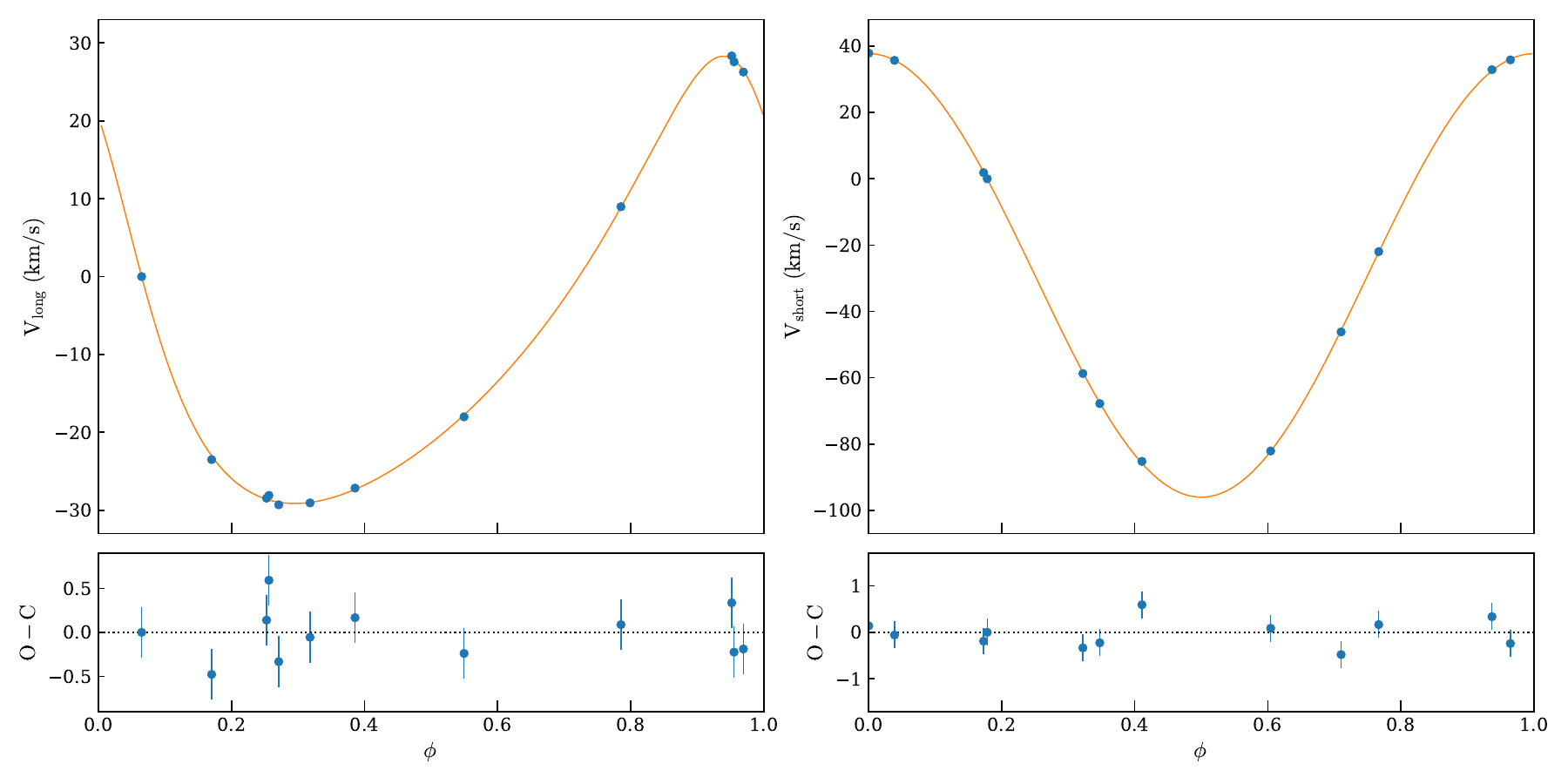}}
	\caption{Fit of the long and short orbital period of the STIS radial velocities.}
	\label{figure__disentangled_RV}
\end{figure}

As first guess parameters, we used the values given in \citet{Evans_1990_06_0}. We display in Fig.~\ref{figure__disentangled_RV} the disentangled long- and short-period orbital velocities and the fitted parameters in Table~\ref{table__fitted_short_and_long_RV}. We used a standard least-squares minimization with errors taken from the correlation matrix. As the STIS uncertainties are only internal (mean standard deviation of segments for all the spectra), we rescaled the RVs uncertainties to the mean r.m.s. of the fit, which was of $\sim0.29$\,km~s$^{-1}$. Our measured parameters are in agreement with the one estimated by  \citet{Evans_1990_06_0}, who performed a similar analysis using high-dispersion spectrographs from the IUE satellite. Additionally, the small residuals for the short-period orbit support our hypothesis of a circular orbit.

We then analytically removed the short-period motion from the STIS velocities which will be used in our combined fit of Sect.~\ref{section__orbital_fitting}.

	\begin{table}[!h]
	\centering
	\caption{Best fit parameters of the STIS differential velocities of the long- and short-period orbit.}
	\begin{tabular}{cc}
		\hline
		\hline
		\multicolumn{2}{c}{Long orbit} \\
		$P_\mathrm{L}$ (days) & $548.0\pm1.2$   \\
		$T_\mathrm{p,L}$ (days) & $2456943.2\pm2.2$  \\
		$e$ & $0.324\pm0.006$   \\
		$K_\mathrm{B}$ (km~s$^{-1}$) & $28.72\pm0.23$   \\
		$\omega$ & $223.18\pm2.12$   \\
		$a_\mathrm{B} \sin i$ (au) & $1.37\pm0.01$   \\
		$f(M_\mathrm{A}, M_\mathrm{B})$ (M$_\odot$) & $1.14\pm0.03$   \\
		\hline
		\multicolumn{2}{c}{Short orbit} \\
		$P_\mathrm{S}$ (days) & $4.67529\pm0.00001$   \\
		$T_\mathrm{p,S}$ (days) & $2456977.994\pm0.003$   \\
		$K_\mathrm{Ba}$ (km~s$^{-1}$)  & $66.89\pm0.21$   \\
		$a_\mathrm{Ba} \sin i$ (au) & $0.0287\pm0.0001$   \\
		$f(M_\mathrm{Ba}, M_\mathrm{Bb})$ (M$_\odot$) & $0.145\pm0.001$   \\
		\hline
	\end{tabular}
	\label{table__fitted_short_and_long_RV}
\end{table}

\subsection{Radial velocities from the literature}

\begin{figure}[!h]
	\centering
	\resizebox{\hsize}{!}{\includegraphics{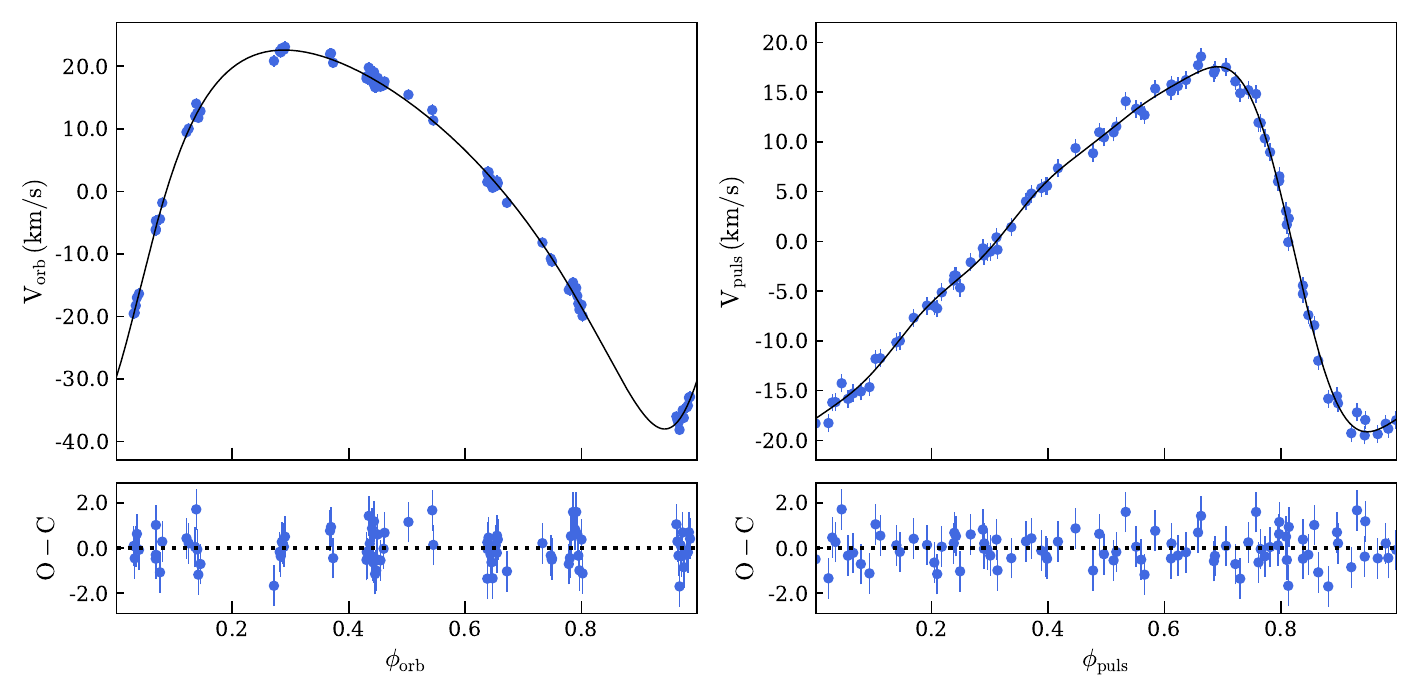}}
	\resizebox{\hsize}{!}{\includegraphics{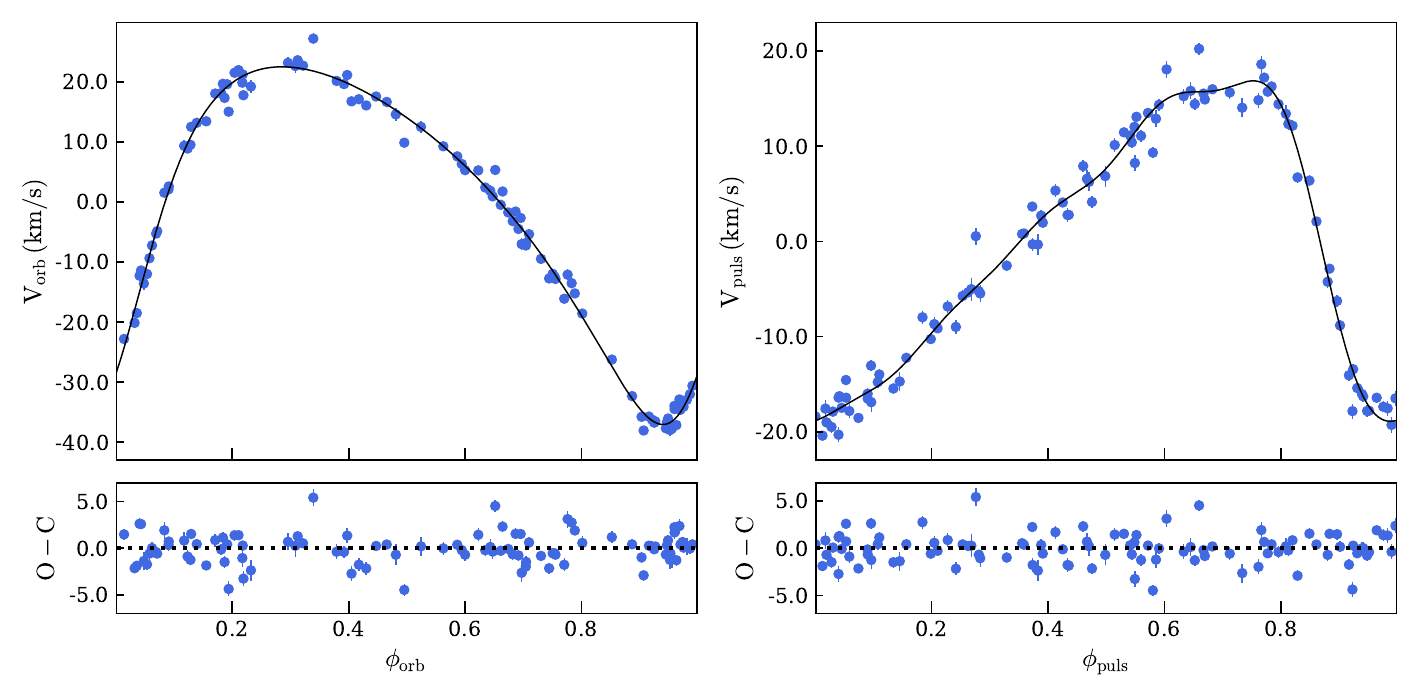}}
	\caption{Fit of the orbital and pulsation motion of the Cepheid using data from Imbert (top panels) and Evans (bottom panels). The fitted model is represented with a black line and the observations as blue dots.}
	\label{figure__RV_cepheid}
\end{figure}

In the visible wavelength, where most of the ground-based spectrograph operate, the Cepheid dominates the spectra. Therefore, RVs obtain from these observations give us information about the orbital reflex motion of the Cepheid. There are several dataset available in the literature, but some suffered from large uncertainties or large scatter in the measurements so we decided to not include them in the fit, which are the case of RVs from \citep{Hellerich_1919_11_0,Barnes_1987_10_1,Wilson_1989_04_0}. The velocities from \citet{Borgniet_2019_11_0} are of very good quality, unfortunately there are only 13 measurements with a small orbital coverage. There are a large set of observations from \citet{Evans_1989_06_0} spanning about six years, providing a good coverage of the long orbital period. However, the mean scatter of the data are on the order of 1.6\,km~s$^{-1}$ when fitting the orbital and pulsation motion. This is likely due to a number of changes made to the spectrograph during those six years which might result in some residual offsets between datasets, although some corrections were done by the authors. The method used to determine the RVs may also contribute to the scatter as they fitted parabola to the line cores of about 20 selected lines, which is less robust than the cross-correlation technique. \citet{Imbert_1984_12_0} obtained spectra from the CORAVEL instrument \citep{Mayor_1983_12_0} spanning about three years, i.e. about two times the long orbital period. The RVs were extracted using the cross-correlation technique, providing more accurate measurements. We found a mean scatter $\sim0.6$\,km~s$^{-1}$ when fitting the orbital and pulsation motion. A comparison of Imbert and Evans dataset is displayed in Fig.~\ref{figure__RV_cepheid}. We preferred to use a uniform more accurate dataset, therefore we only used the measurements from  \citet{Imbert_1984_12_0} for our simultaneous fit of the next section.

\subsection{Astrometry}
 \label{subsection__astrometry}
  
 To detect the companion (or to be more precise the centre of light in our specific case of a binary companion), we used the interferometric tool \emph{CANDID}\footnote{Available at \url{https://github.com/amerand/CANDID} and \url{https://github.com/agallenne/GUIcandid} for a GUI version.} \citep{Gallenne_2015_07_0}. The main function allows a systematic search for companions performing an $N\times N$ grid of fits, the minimum required grid resolution of which is estimated a posteriori in order to find the global minimum in $\chi^2$. The tool delivers the binary parameters, namely the flux ratio $f$, and the relative astrometric separation ($\Delta \alpha, \Delta \delta$), together with the uniform-disk angular diameter $\theta_{UD}$ of the primary star (the Cepheid). The angular diameter of the companion is assumed to be unresolved by the interferometer. The significance of the detection, $n\sigma$, is also given, taking into account the reduced $\chi^2$ and the number of degrees of freedom\footnote{with a maximum displayed of $8\sigma$, because $p$-values are converted into chi-squared statistics, but for values very close to 1 the computer precision is reached.}. They are listed in Table~\ref{table__astrometry_results}, together with our measured astrometric positions.
 
 For the 2016 data, we had poor seeing conditions, we therefore combined the two nights to detect the companion and used only the closure phase signal which is less sensitive to the atmospheric variations. We fixed the angular diameter to the average diameter measured from all our observations (see next paragraph). The night of 2020-06-30 also suffered from poor observing conditions, so we decided to combine with the following night to increase the S/N, again using only the closure phases. We fixed the angular diameter to the average value given by fitting the $V^2$. The data taken during two nights in 2022 suffered from vibrations on the delay lines systems, which affects mostly the visibilities as the order of magnitude is less than a wavelength. We therefore also only used the closure phase with a fixed flux ratio. To increase the detection level, we combine these two nights for MIRC-X and MYSTIC. Both instruments provide identical detection of the companion (see Table~\ref{table__astrometry_results}).
 
 Uncertainties on the fitted parameters are estimated using a bootstrapping function (bootstrap on the observing time and baselines). From the distribution, we took the median value and the maximum value between the 16th and 84th percentiles as uncertainty for the flux ratio and angular diameter. For the fitted astrometric position, the error ellipse is derived from the bootstrap sample (using a principal components analysis).The angular diameters and astrometric separations were then divided by factors of $1.0014 \pm 0.0006$ for MIRC, $1.0054 \pm 0.0006$ for MIRC-X and $1.0067 \pm 0.0007$ for MYSTIC (J.D. Monnier, private communication) to take into account the uncertainty from the wavelength calibration. This is equivalent to adjusting the respective wavelengths reported in the OIFITS files by the same factors.
 
 Uncertainty on the angular diameter measurements was estimated using the conservative formalism of  \citet{Boffin_2014_04_0} as follows:
 \begin{displaymath}
	\sigma^2_\mathrm{\theta_{UD}} = N_\mathrm{sp} \sigma^2_\mathrm{stat} + \delta \lambda^2 \theta^2_\mathrm{\theta_{UD}}
 \end{displaymath}
where $N_\mathrm{sp}$ is the number of spectral channels, $\sigma^2_\mathrm{stat}$ the uncertainty from the bootstrapping and $\delta \lambda = 0.07$\,\%, as mentioned above. We measured a mean uniform disk diameter in the $H$ band of $\mathrm{\theta_{UD}} = 0.323\pm0.061$\,mas (the standard deviation is taken as uncertainty), which is in agreement with the value of \citet{Trahin_2019_11_0} estimated from a global fit. We also estimated an average flux ratio in $H$ of $f_\mathrm{H} = 1.68\pm0.24$\,\%.
 
	\begin{table*}[!h]
	\centering
	\caption{Relative astrometric position of the SU~Cyg companion.}
	\begin{tabular}{cccccccccc}
		\hline
		\hline
		Date  & HJD  &  $\Delta \alpha$	&	$\Delta \delta$	& $\sigma_\mathrm{PA}$  & $\sigma_\mathrm{maj}$& $\sigma_\mathrm{min}$  & $f$  & $\theta_\mathrm{UD}$ & $n\sigma$ \\
 				   &   (Day) & (mas) & (mas) & (deg) & (mas) & (mas) & ($\%$) & (mas) & \\
		\hline
		2016-07-19\tablefootmark{b} 	& 2457588.299 & -2.482 & -0.369 & 70.3 & 0.080 & 0.062  & $1.24\pm0.19$ & 0.323 & 5.3 \\
		2019-07-04 									& 2458678.779 & -2.084 & -0.434 & 99.9 & 0.020 & 0.016  & $1.62\pm0.11$ & $0.332\pm0.013$ & $>8$ \\
		2020-07-01\tablefootmark{c}	 & 2459031.313 & 1.135 & 0.533 & 144.8 & 0.015 & 0.007   & $1.66\pm0.07$ & $0.231\pm0.020$ & $>8$ \\
		2021-07-20 									& 2459415.768 & -2.975 & 0.219 & -160.4 & 0.025 & 0.012  & $2.03\pm0.16$ & $0.328\pm0.026$ & $>8$ \\
		2021-09-20 									& 2459476.675 & -1.799 & 0.426 & 145.7 & 0.044 & 0.024  & $1.80\pm0.26$ & $0.400\pm0.034$ & $>8$ \\
		2022-08-21\tablefootmark{d} & 2459812.622 & -3.049 & -0.323 & 109.8 & 0.041 & 0.017  & $1.75\pm0.11$ & 0.323 & $>8$ \\
		2022-08-21\tablefootmark{e} & 2459812.653 & -3.087 & -0.310 & 149.8 & 0.021 & 0.011  & $1.74\pm0.07$\tablefootmark{a} & 0.323 & $>8$ \\
		\hline
	\end{tabular}
	\label{table__astrometry_results}
\tablefoot{\tablefoottext{a}{Flux ratio mentioned here is in $K$ because of MYSTIC observations, while others are in $H$.}
				\tablefoottext{b}{The nights of July 18 and 19 were combine as mentioned in the text.}
				\tablefoottext{c}{The nights of June 30 and July 01 were combine  as mentioned in the text.}
				\tablefoottext{d}{MIRCX observations. The nights of August 20 and 21 were combine  as mentioned in the text.}
				\tablefoottext{e}{MYSTIC observations. The nights of August 20 and 21 were  combine  as mentioned in the text.}
}
\end{table*}
 
As the companion is itself a binary, the astrometry we measured is the one of the photocentre of the companion pair. For stars with equal brightness, the photocentre would be in the middle of the two stars and would follow a simple elliptical orbit around the Cepheid. However, this is not our case because $Ba$ is a B7.5V star and $Bb$ has a spectral type later than A0V, which means that the photocentre has a wobble around its elliptical orbit around the Cepheid. The amplitude of the wobble can be roughly quantified from the mass ratio and contrast between the stars. Using the spectral type calibration from \citet{Pecaut_2013_09_0}, the companions have a mass close to $3.6\,M_\odot$ and $\leq 2.2\,M_\odot$, respectively for $Ba$ and $Bb$, and a magnitude difference in $\Delta H$ of $\sim1.1$\,mag. The semimajor axis $a_\mathrm{ph}$ of the photocentre orbit of the companion pair can be estimated with:
\begin{displaymath}
	a_\mathrm{ph} = a_\mathrm{short} \left( \dfrac{M_{Bb}}{M_{Ba}+M_{Bb}} - \dfrac{1}{1 + 10^{0.4*\Delta H}} \right)
\end{displaymath}
with $a_\mathrm{short}$ the semimajor axis of the short orbit in arcsecond, and $M_{Ba}$ and $M_{Bb}$ the mass of each component. $a_\mathrm{short}$ can be also estimated from the centre of mass equation:
\begin{displaymath}
	a_\mathrm{short} = a_{Ba} + a_{Bb} = a_{Ba} \left( 1+\dfrac{M_{Ba}}{M_{Bb}} \right)
\end{displaymath}
where $a_{Ba}$ and $a_{Bb}$ are the distance between the centre of mass and the star $Ba$ and $Bb$, respectively. Using $a_{Ba}$ from Table~\ref{table__fitted_short_and_long_RV}, the inclination of 80$^\circ$ (estimated a posteriori, see Sect.~\ref{section__orbital_fitting}), the Gaia distance of 1000\,pc and the predicted masses, we estimated $a_\mathrm{short} \sim 0.1$\,mas, and a semimajor axis of the photocentre of $a_\mathrm{ph} \sim 11\,\mu$as. This is slightly below our average astrometric precision. To be conservative, we added to our astrometry an additional uncertainty of $15\,\mu$as.
 
\section{Orbital fitting}
\label{section__orbital_fitting}
 
We performed a combined fit of the interferometric orbit and RVs of the Cepheid and its companion pair, i.e. of the long period orbit. This is similar to the work we did in \citet{Gallenne_2018_11_0} for the Cepheid \object{V1334~Cyg}. The fit includes three models with shared parameters for our three data sets. The first models the Cepheid RVs only, i.e., the orbit around the system barycentre and the pulsation. The second models the STIS radial velocities, i.e., the orbit of the centre of mass of the companion pair around the system barycentre with the Cepheid. The last model fits the relative astrometric orbit of the companion pair determined from interferometry (MIRC data).

Our radial velocity model of the primary star, the Cepheid, is defined as
\begin{displaymath}
	V_\mathrm{A} = V_\mathrm{A,orb} + V_\mathrm{puls},
\end{displaymath}
with the orbital radial velocity $V_\mathrm{A,orb}$ and the pulsation velocity $V_\mathrm{puls}$, which are expressed with
\begin{align*}
	V_\mathrm{A,orb} &= K_\mathrm{A} [\cos(\omega + \nu) + e \cos \omega] + \gamma \\
	V_\mathrm{puls} &= \sum_{i=1}^n [A_i \cos(2\pi i \phi_\mathrm{puls}) + B_i \sin(2\pi i \phi_\mathrm{puls})],
\end{align*}
with $K_A$ the semi-amplitude of the Cepheid’s orbit due to the secondary companions, $\omega$ the argument of periastron of the companion pair orbit, $\nu$ the true anomaly of the companions, $e$ the eccentricity of the orbit, $\gamma$ the systemic velocity, the pulsation phase $\phi_\mathrm{puls} = (t - T_0)/P_\mathrm{puls}$ (modulo 1), $T_0$ is the reference epoch (usually defined at the maximum brightness), and $(A_i, B_i)$ the amplitude of the Fourier series. The true anomaly is defined implicitly with the following three Keplerian parameters $P_\mathrm{orb} , T_p$ and $e$ and Kepler's equation:
\begin{align*}
	\tan \frac{\nu}{2} &= \sqrt{\frac{1 + e}{1 - e} } \tan \frac{E}{2}\\
	E - e \sin E &= \frac{2\pi (t - T_p)}{P_\mathrm{orb}},
\end{align*}
where $E(t)$ is the eccentric anomaly, $T_p$ the time of periastron passage, $t$ the time of radial velocity observations, and $P_\mathrm{orb}$ the (long) orbital period.

Following the linear parametrization developed in \citet{Wright_2009_05_0} for $V_\mathrm{A,orb}$, and including now the pulsation, our model can be simplified to
\begin{equation*}
	\begin{split}
		V_\mathrm{A}& = C_1 \cos \nu + C_2 \sin \nu + V_0\\
		&+ \sum_{i=1}^4 [A_i \cos(2\pi i \phi_\mathrm{puls}) + B_i \sin(2\pi i \phi_\mathrm{puls})
	\end{split}
\end{equation*}
where we restricted the Fourier series to $n = 4$. The parameters $C_1, C_2$ and $V_0$ are related to the Keplerian parameters through the relations \citep{Wright_2009_05_0}:
\begin{align*}
	C_1 &= K_\mathrm{A} \cos \omega, \\
	C_2 &= -K_\mathrm{A} \sin \omega, \\
	V_0 &= \gamma + K_\mathrm{A} e \cos \omega,
\end{align*}
which can be converted back with ($\omega$ chosen so that $\sin \omega$ has the sign of the numerator):
\begin{align*}
	K_\mathrm{A} &= \sqrt{C_1^2 + C_2^2}, \\
	\tan \omega &= \frac{-C_2}{C_1},\\
	\gamma &= V_0 - K_\mathrm{A} e \cos \omega.
\end{align*}
The fitted parameters are therefore defined as ($C_1, C_2$, $V_0, A_1, B_1, A_2, B_2, P_\mathrm{orb}, T_p, e, P_\mathrm{puls}$). $T_0$ is kept fixed in the fitting process to avoid degeneracy with $T_p$, and its value is taken from \citet{Trahin_2019_11_0}.


For the second model, we used the short-orbit corrected STIS spectra from Sect.~\ref{subsection__stis_radial_velocities}, which refer to the relative radial velocities of the companion pair (named BC here) in their long orbit with the Cepheid. They can be parametrized with:

\begin{align*}
	\Delta V_\mathrm{B}(t) &= V_\mathrm{B}(t) - V_\mathrm{B}(t_0), \\
	V_\mathrm{B} &= \frac{K_\mathrm{A}}{q} [\cos(\omega + \nu) + e \cos \omega] + \gamma\\
	q &= \dfrac{M_\mathrm{B}}{M_\mathrm{A}} = \dfrac{M_\mathrm{Ba}+M_\mathrm{Bb}}{M_\mathrm{A}}
\end{align*}
with the mass ratio $q$ and $t_0$ the time of the first STIS measurement. This adds the new parameter, $q$ as fitted parameter. We also added an instrument zero point offset $zp$ for the STIS velocities as fitted parameter.
	 
Finally, the astrometric positions of the companion pair as measured from interferometry are modelled with the following equations:
	\begin{align*}
	\Delta \alpha &=   r \,[ \cos \Omega \cos(\omega + \nu) - \cos i \sin \Omega \sin(\omega + \nu) ], \\
	\Delta \delta &= r \,[ \sin \Omega \cos(\omega + \nu) + \cos i \cos \Omega \sin(\omega + \nu) ], 
\end{align*}
with $a$ the angular semi-major axis in arcsecond, $\Omega$ the position angle of the ascending node, and $i$ the orbital inclination. The true anomaly $\nu$ and the separation $r$ at a given time $t$ are calculated as:
\begin{displaymath}
	r = a (1 - e \cos E),
\end{displaymath}
where the eccentric anomaly $E$ is calculated by solving the Kepler's equation as previously.

At each iteration, the times of observations of the RV and interferometric observations were adjusted  to account for the LTTE in the long orbit.

The distance and masses are then given by:
\begin{align*}
	K_\mathrm{B} &= \frac{K_\mathrm{A}}{q}, \\
	a_\mathrm{au} &=  \frac{9.191966\times10^{-5} (K_\mathrm{A} + K_\mathrm{B} ) P_\mathrm{orb} \sqrt{1 - e^2}}{\sin^3 i}, \\
	d &= \frac{a_\mathrm{au}}{a_\mathrm{as}}, \\
	M_T &= M_\mathrm{A} + M_\mathrm{B}  = \frac{a^3 d^3}{P_\mathrm{orb}^2}, \\
	M_\mathrm{A} &= \frac{M_T}{1 + q}, \\
	M_\mathrm{B}  &= q\,M_\mathrm{A},
\end{align*}
with $K_\mathrm{B} $ the semi-amplitude of the companion pair, $a_\mathrm{au}$ the linear semi-major axis in astronomical units (au), $d$ the distance to the system, and $M_\mathrm{A}$ and $M_\mathrm{B}$ the Cepheid and the masses of the two companions, respectively.

The combined fit is performed using a Monte Carlo Markov Chain (MCMC) technique\footnote{With the Python package \emph{emcee} developed by \citet{Foreman-Mackey_2013_03_0}} to fit all model parameters characterizing the standard orbital elements and the pulsation of the Cepheid. As a starting point for our 100 MCMC walkers, we performed a least squares fit using spectroscopic orbital values from \citet{Evans_1990_06_0} as first guesses, while $a, i$ and $\Omega$ were set to 3\,mas, 50$^\circ$ and 50$^\circ$, respectively. We then ran 100 initialization steps to well explore the parameter space and get settled into a stationary distribution. For all cases, the chain converged before 50 steps. Finally, we used the last position of the walkers to generate our full production run of 1000 steps, discarding the initial 50 steps. All the orbital elements are estimated from the distribution taking the median value and the maximum value between the 16th and 84th percentiles uncertainty (although the distributions were roughly symmetrical). To be more conservative, we also quadratically added a 0.25\,\% error to the distance uncertainty due to the wavelength calibration of the MIRC instrument (although it is likely $< 0. 25$\,\% as mentioned in Sect.~\ref{subsection__astrometry}).

\begin{figure*}[ht]
	\centering
	\resizebox{\hsize}{!}{\includegraphics{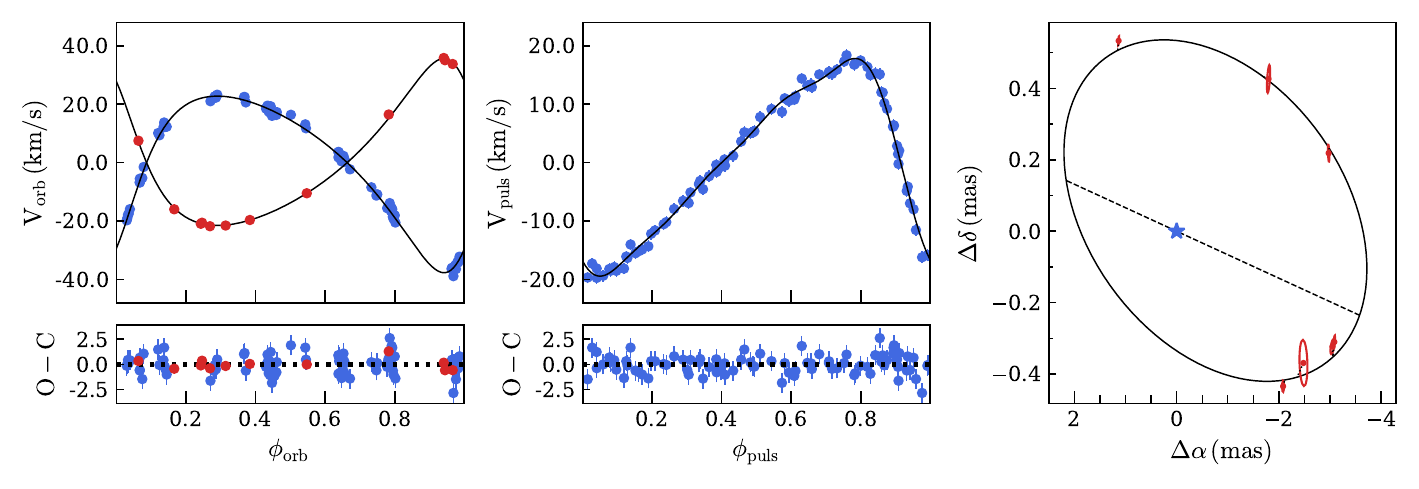}}
	\caption{Result of our combined fit. Left: fitted (solid lines) and measured primary (blue dots) and secondary (red dots) orbital velocity. Middle: fitted (solid line) and extracted (blue dots) pulsation velocity. Right: relative astrometric orbit of the companion pair.}
	\label{figure_orbit}
\end{figure*}

\begin{figure*}[ht]
	\centering
	\resizebox{\hsize}{!}{\includegraphics{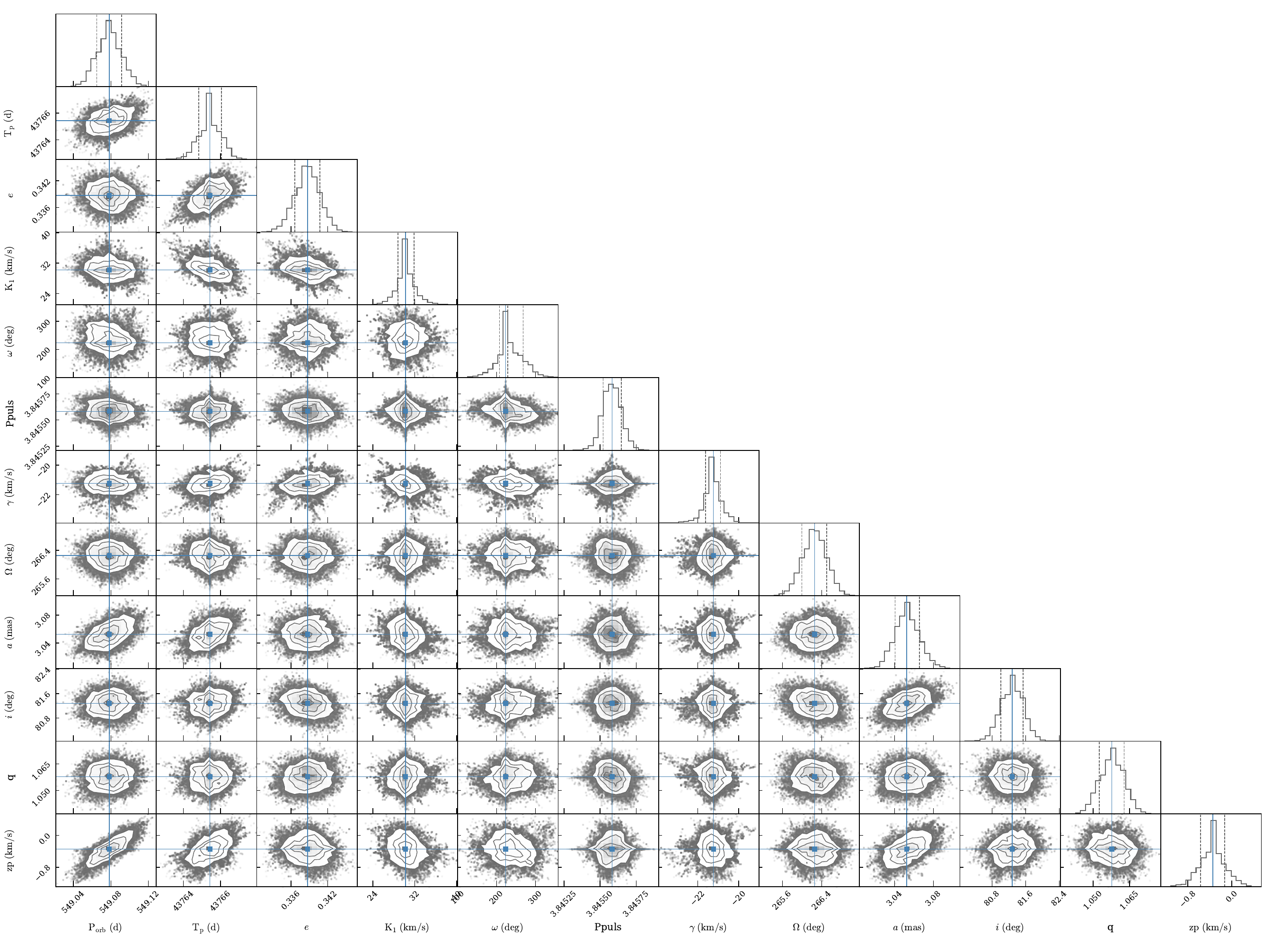}}
	\caption{Corner plots of the Markov chain Monte Carlo posterior distributions produced from \emph{emcee} for the long period orbit.}
	\label{figure_cornerplot}
\end{figure*}

Our best fit is shown in Fig.~\ref{figure_orbit}. We determined a distance $d = \dist \pm \edist$\,pc ($\pm \accuracydist$\,\%), which is the most accurate model-independent distance of a Cepheid, slightly more accurate than the previous measurement of V1334~Cyg distance \citep{Gallenne_2018_11_0}. This corresponds to a parallax of $\varpi = \parallax \pm \eparallax$\,mas. The masses were also determined with a high accuracy: $M_\mathrm{A} = \Mcep \pm \eMcep\,M_\odot$ (\accuracyMcep\,\%) and $M_\mathrm{B}  = \Mcomp \pm \eMcomp\,M_\odot$ (\accuracyMcomp\,\%), with the primary star being the Cepheid. The other derived orbital and pulsation parameters are listed in Table~\ref{table_orbit}, with our posterior distribution from our MCMC analysis displayed in Fig.~\ref{figure_cornerplot}. We found a zero point offset $zp = -0.35 \pm 0.23\mathrm{km~s^{-1}}$ for the STIS velocities.

		\begin{table}[!h]
	\centering
	\caption{Final estimated parameters of the SU~Cyg system.}
	\begin{tabular}{cc} 
		\hline
		\hline
		\multicolumn{2}{c}{Pulsation}																	\\
		$P_\mathrm{puls}$ (days)							  & $3.84559 \pm 0.00007$\\
		$T_0$ (JD)						   						 	 & $2~433~301.777$		\\
		$A_1$ (km~s$^{-1}$)					&	$ -8.99 \pm 0.43$			\\
		$B_1$ (km~s$^{-1}$)					&	$-13.59 \pm 0.24$	\\
		$A_2$ (km~s$^{-1}$)					&	$-4.56 \pm 0.21$		\\
		$B_2$ (km~s$^{-1}$)					&	$-2.86 \pm 0.25$		\\
		$A_3$ (km~s$^{-1}$)					&	$-2.39 \pm 0.08$		\\
		$B_3$ (km~s$^{-1}$)					&	$-0.24 \pm 0.24 $		\\
		$A_4$ (km~s$^{-1}$)					&	$-0.81 \pm 0.06$		\\
		$B_4$ (km~s$^{-1}$)					&	$0.56 \pm 0.16$		\\
		\hline
		\multicolumn{2}{c}{Orbit}  																			\\
		$P_\mathrm{orb}$ (days)									&   $549.077 \pm0.013$ 		  \\
		$T_\mathrm{p}$ (JD)										&  $2443765.94 \pm0.63$   \\
		$e$																     &  $0.339 \pm 0.002$	\\
		$\omega$	($^\circ$)									 &  	$223.16 \pm0.42$		  \\
		$K_{A}$ ($\mathrm{km~s^{-1}}$)					  & 	$30.25 \pm0.05$	\\
		$K_\mathrm{B}$ ($\mathrm{km~s^{-1}}$)					  &		$28.59 \pm0.15$	\\
		$\gamma$	($\mathrm{km~s^{-1}}$)			&  	$-21.23 \pm0.06$	\\
		$\Omega$	($^\circ$)									 &		$266.24 \pm0.24 $	\\
		$i$ ($^\circ$)													&		$81.28 \pm 0.27$	\\
		$a$ (mas)														 &		$3.052 \pm 0.013$	\\
		$a$ (au)														 &		$2.827 \pm0.009$	 \\
		$d$ (pc)														 &		$\dist \pm \edist$	\\
		$\varpi$ (mas)														 &		$\parallax \pm \eparallax$	\\
		$q$ 																&		$\Mratio \pm \eMratio$	\\
		$M_\mathrm{A}$ ($M_\odot$)											&		$\Mcep \pm \eMcep$	\\
		$M_\mathrm{B}$ ($M_\odot$)											&		$\Mcomp \pm \eMcomp$	\\
		\hline															
	\end{tabular}
	\label{table_orbit}
	\tablefoot{Index A designates the Cepheid and index B the companion pair. Note that $T_0$ is kept fixed.}
\end{table}

\section{Discussion}
\label{section__discussion}

\subsection{The Cepheid distance}

\begin{figure}[ht]
	\centering
	\resizebox{\hsize}{!}{\includegraphics{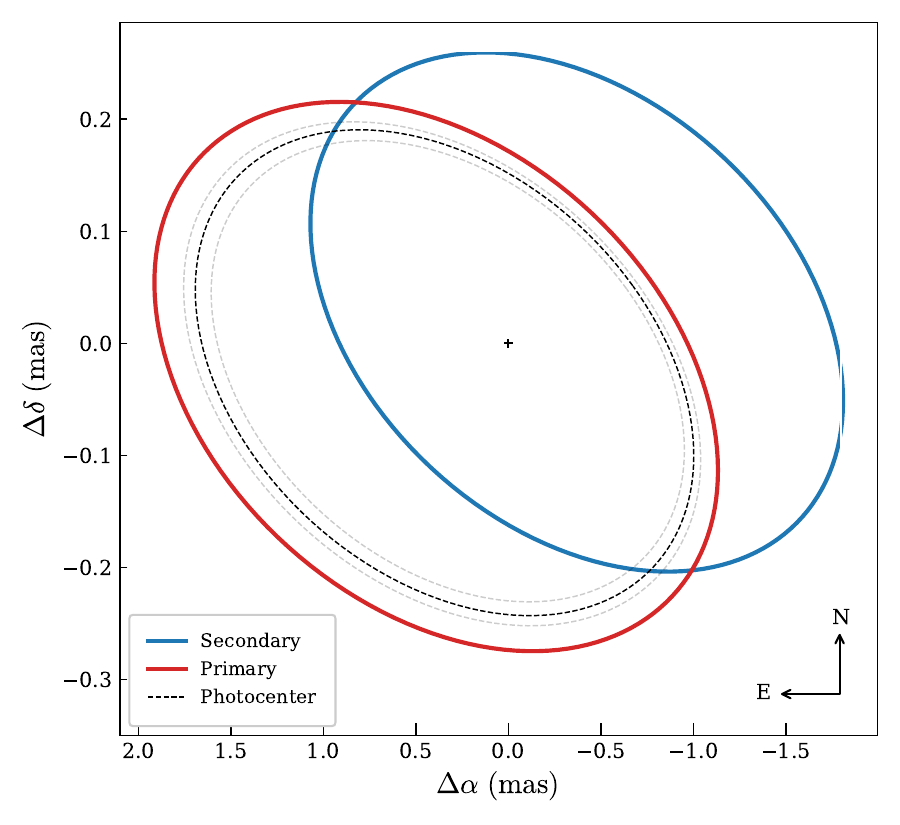}}
	\caption{Orbits of SU Cyg A (blue) and its companion B (red) around their common centre of mass. The average virtual orbit of the photocentre of the system is shown as a dashed black ellipse. The grey ellipses represent the maximum and minimum shifts in the photocenter relative to the average position, caused by the brightness variations of the Cepheid during its pulsation cycle.}
	\label{figure_orbit_photocenter}
\end{figure}

This is the second Cepheid with the most precise and accurate orbital parallax measured to 1\,\% using a combination of interferometry and space- and ground-based spectroscopy. In \citet{Gallenne_2018_11_0}, we measured the orbital parallax of the Cepheid V1334~Cyg with an accuracy of 1\,\% using the same method and showed the disagreement with the most used P-L relations and a deviation of $3.6\sigma$ with the Gaia DR2. Here we will perform a similar analysis with the Cepheid SU~Cyg. Gaia DR3 provides a parallax of $1.000\pm0.052$\,mas, including a zero point correction of $-0.028$\,mas \citep{Lindegren_2021_05_7}, which is in agreement at $1.5\sigma$ with our measurements, but our uncertainty is $\sim9\times$ better. Because the DR3 parallax has large uncertainties, the zero-point correction has no effect on the agreement with our distance estimate. The Renormalized Unit Weight Error (RUWE) indicator of the Gaia astrometric fit is 3.44, meaning that the astrometric solution is problematic, likely due to the orbital motion of the Cepheid. \citet{Wahlgren_1998_04_0} identified the companion B as a HgMn type star of B8V spectral type, giving a magnitude difference with the Cepheid of $\Delta V \sim 3$\,mag \citep{Evans_1995_05_0}. Using our measured orbital parameters and this contrast, we calculated the orbit of the photocentre, which we display in Fig.~\ref{figure_orbit_photocenter}. We see that its semi-major axis is $\sim 1.4$\,mas and certainly impacts the Gaia single-star model solution. Since the Cepheid is redder than its companion, the magnitude difference in the G band is slightly larger than the estimate in the V band, of the order of +0.1-0.2mag, leading us to expect a similar bias in the G-band parallax. In addition, the orbital period is close to one year. Unfortunately, there are no information in the non-single stars catalogue of Gaia for this system \citep{Halbwachs_2023_06_0}. In addition, the Cepheid is in the bright-star magnitude range of Gaia, suffering from saturation effects in the detector which alter the astrometric performances of Gaia.

\begin{figure*}[!ht]
	\centering
	\resizebox{0.9\hsize}{!}{\includegraphics{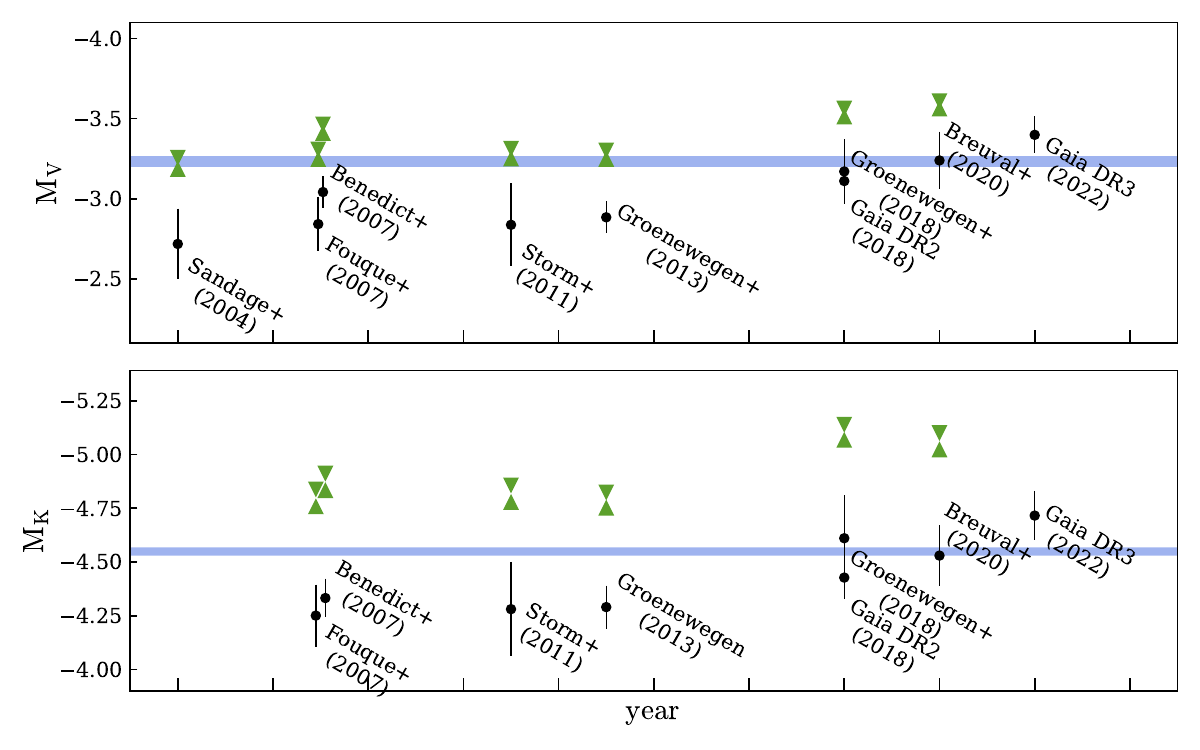}}
	\caption{Comparison between the absolute magnitudes of SU~Cyg predicted from literature P-L relations (black dots) and the present distance measurement (blue areas), in two photometric bands. The blue areas represent the measured absolute magnitude after correcting for the flux contamination from the companion. The up and down green triangles represent the absolute magnitudes determined from literature P-L relations by using a "fundamentalized" pulsation period, with the down symbol for the smallest converted period (5.455\,d) and the up symbol for the longest (5.757\,d). We see that only modern-era P-L calibrations (including Gaia) seem to predict correctly the luminosity of SU Cyg. }
	\label{figure_pl}
\end{figure*}

\begin{figure}[!ht]
	\centering
	\resizebox{\hsize}{!}{\includegraphics{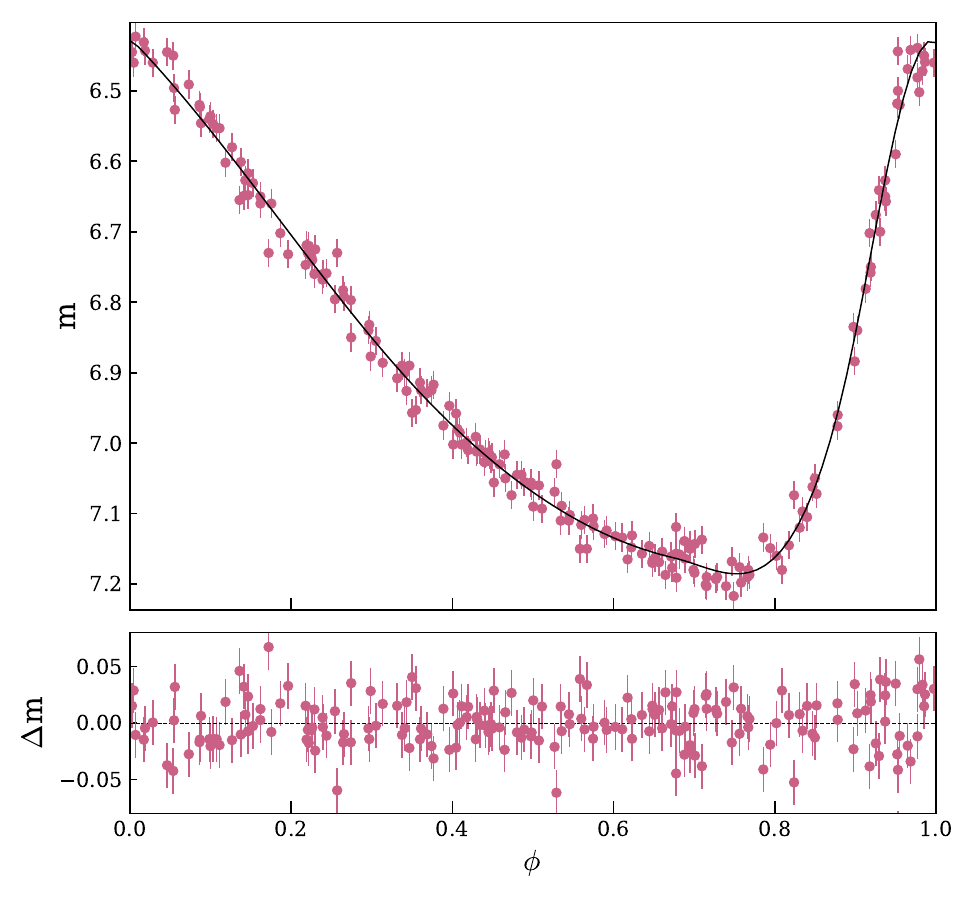}}
	\caption{Fitted $V$-band light curve of SU~Cyg.}
	\label{figure_lightcurve}
\end{figure}

Our precise and independent distance measurement also provide an independent check of the P-L relations for fundamental-mode Cepheids as SU~Cyg is classified as pulsating in its fundamental mode \citep[see e.g.][]{Clementini_2019_02_0,Luck_2018_10_0}. A reliable method to determine whether a star is in the fundamental or first overtone mode is to examine the period-radius relation, which is notably precise. According to the empirical relation from \citet{Gallenne_2017_11_0}, the prediction for a fundamental mode is $\sim 34\,R_\odot$, aligning well with our estimate of $\Rcep\pm\eRcep\,R_\odot$. Thus, the Cepheid is consistent with being a fundamental mode pulsator. We selected a sample of published relations in the $V$- and $K$-band filters, as they are the most frequently used photometric bands \citep{Sandage_2004_09_0,Fouque_2007_12_0,Benedict_2007_04_0, Storm_2011_10_0,Groenewegen_2013_02_0,Groenewegen_2018_11_0,Breuval_2020_11_0}. For each tested P-L relation, we determined the predicted absolute magnitudes for $P_0$, that are represented in Fig.~\ref{figure_pl} as black dots. We adopted as uncertainties the scatter of each relation given by the authors, which better represents the observed intrinsic dispersion (the width of the instability strip is the dominating uncertainty in the P-L relations). We derived the absolute magnitude with the usual relation $M_\lambda = m_\lambda - A_\lambda - 5\,\log d + 5$, with $m_\lambda$ the apparent magnitude at wavelength $\lambda$, $A_\lambda$ the interstellar absorption parameter, and $d$ the distance to the star. We corrected for the interstellar extinction, using $A_\mathrm{V} = 3.23 E(B-V)$ and $A_\mathrm{K} = 0.119 A_\mathrm{V}$ \citep{Fouque_2007_12_0}, with $E(B-V) = 0.109 \pm 0.006$ \citep{Trahin_2021_12_5}. For the $V$ band, we estimated the weighted mean apparent magnitude using a periodic cubic spline fit of four different published light curves \citep{Berdnikov_2008_04_0,Kiss_1998_07_0,Moffett_1984_07_0,Szabados_1977_01_0}. The curve is shown in Fig.~\ref{figure_lightcurve}. We determined the weighted mean dereddened magnitude $m_\mathrm{0,V} = 6.602 \pm 0.024$\,mag, where the total uncertainty is estimated from the standard deviation of the residual values. In the near-infrared, we used the mean apparent magnitude $m_\mathrm{K} = 5.307 \pm 0.015$\,mag from \citet{Monson_2011_03_0}, determined from near-IR light curve. We estimated the dereddened magnitude $m_\mathrm{0,K} = 5.285 \pm 0.015$\,mag.

We then combined with our measured distance to estimate the absolute magnitudes $M_\lambda$, for which the total error bar includes the uncertainties on $d, A_\lambda$, and $m_\lambda$. Fig.~\ref{figure_pl} shows the difference between $M_\lambda$ of SU~Cyg (red area) and the predicted values from literature P-L relations (black dots). However, the contribution of the companion must be subtracted from the magnitudes. To correct from the flux contamination of the companion, we used the average magnitude difference between the components estimated in $V$ by \cite{Evans_1995_05_0}, $\Delta V = 3$\,mag, and from our measured $H$-band average flux ratio from interferometry for the $K$ band, i.e. we assumed $\Delta K = \Delta H = 4.44 \pm 0.16$\,mag. With our distance we can estimate the mean corrected absolute magnitudes of the Cepheid to be $M\mathrm{_V(cep)} = -3.23 \pm 0.03$\,mag, $M\mathrm{_H(cep)} = -4.50 \pm 0.02$\,mag and $M\mathrm{_K(cep)} = -4.55\pm 0.02$\,mag. Note that we assumed a similar flux ratio in $H$ and $K$, but this a a good approximation as the companion is rather faint in the near-IR and our MIRCX and MYSTIC observations provide similar flux ratios in $H$ and $K$ band, respectively (see Table~\ref{table__astrometry_results}). From the magnitude differences, absolute magnitudes for the companion pair can also be derived, we found $M\mathrm{_V(B)} = -0.22 \pm 0.11$\,mag, $M\mathrm{_H(B)} = -0.07 \pm 0.16$\,mag and $M\mathrm{_K(B)} = -0.11 \pm 0.16$\,mag. The spectral type of B8V identified by \cite{Evans_1995_05_0} for the companion Ba corresponds to an absolute magnitude $M\mathrm{_V(Ba)} \sim 0.0$\,mag \citep{Pecaut_2013_09_0}\footnote{see also \url{http://www.pas.rochester.edu/~emamajek/EEM_dwarf_UBVIJHK_colors_Teff.txt}}, which when combined with $M\mathrm{_V(B)}$ gives a flux ratio of $f_\mathrm{Bb}/f_\mathrm{Ba} \sim 22$\,\% in $V$, and so an absolute magnitude for the component Bb of $M\mathrm{_V(Bb)} \sim 1.6$\,mag. This would correspond to a A3V-A4V spectral type \citep{Pecaut_2013_09_0}, in agreement with the suggestion of \cite{Evans_1995_05_0}. The expected mass from such spectral types is about $3.4\,M_\odot$ and $1.9\,M_\odot$, respectively for Ba and Bb.

The blue area in Fig.~\ref{figure_pl} shows the absolute magnitudes of the Cepheid corrected from the flux contamination. We see that only the relations based on direct distance estimates are in acceptable agreement ($< 1.4\sigma$) with our measured corrected absolute magnitude of the Cepheid. This small discrepancy for Gaia can be explained by the fact that the measurements are still not accurate and suffer from various effect for pulsating stars (chromaticity correction, saturation, etc.), in addition to the binarity effect which is not taken into account yet. The relation from \citet{Benedict_2007_04_0} based on HST Fine Guidance Sensor parallax is at $1.8\sigma$ and $2.4\sigma$, respectively in $V$ and $K$. \citet{Breuval_2020_11_0} shows a very good agreement, where we used the Gaia DR2 parallaxes of wide companions of Cepheids or their host open cluster as proxy of the Cepheid distances, which are thought to be less prone of systematics. The other relations based on photometric measurements, calibrated using the Baade-Wesselink method \citep{Baade_1926_11_0,Wesselink_1946_01_0} to determine the distance of Cepheids, are in agreement at $1.5-3.3\sigma$ in $V$ and $1.2-2.5\sigma$ in $K$. 

SU~Cyg seems brighter than expected, by $\sim 0.15$\,mag in $V$ and $\sim 0.08$\,mag in $K$. One explanation would be the presence of a circumstellar envelope (CSE), which we know exist around other Cepheids \citep{Kervella_2006_03_0,Merand_2006_07_0,Gallenne_2012_02_0,Gallenne_2013_10_0,Hocde_2020_09_0,Hocde_2021_07_9,Gallenne_2021_07_3}. The infrared excess is similar to \citet{Hocde_2020_01_0} who estimated for this Cepheid an excess of $\sim0.04$\,mag using a multi-wavelength  global analysis with the SPIPS code \citep[SpectroPhoto-Interferometry of Pulsating Stars][]{Merand_2015_12_0} and a simple power-law parametrization. However, this model assumes dust emission and no excess for $\lambda < 1.2\,\mu$m. In \citet{Hocde_2020_01_0}, we also modelled the CSE emission with a shell of ionized gas to allow a deficit or excess in the visible. However, this model showed a deficit in the visible, i.e. an absorption instead of an emission, no excess in $K$. Therefore, no CSE models could explain the possible offset to reconcile the photometrically-based P-L relations.

Another explanation would be a wrong estimate of the pulsation period, in the sense that the Cepheid does not pulsate in its fundamental mode but instead in its first overtone mode. To test this hypothesis, we can assume that the pulsation period we measure is the first overtone and we can convert it to a fundamental mode. Several works exit on this topic for Milky-Way and Magellanic Clouds Cepheids but there is still no consensus about a good empirical or theoretical relation to "fundamentalize" the first overtone period \citep{Alcock_1995_04_0,Feast_1997_03_0,Kovtyukh_2016_08_0,Sziladi_2018_06_0,Pilecki_2021_04_0,Pilecki_2024_06_0}. The most used is from \citep{Alcock_1995_04_0} using LMC Cepheids, who estimated the ratio of the first overtone to the fundamental period as $P_1/P_0 = 0.720 - 0.027 \log P_0$, while the latest relation is $P_0 = P_1 (1.367 + 0.079 \log P_1)$ from \citet{Pilecki_2024_07_0} for MW Cepheids. All published relations provides a "fundamentalized" period for SU~Cyg between 5.434 and 5.757\,days, but they do not provide either a consistent agreement with our measured value. We can see in Fig.~\ref{figure_pl} the recalculated values of the P-L relations using 5.434\,days and 5.757\,days, represented by down and up triangles, respectively. In $V$, the photometric-based relation would be in agreement, however, the relations based on the Gaia parallaxes are now not consistent. In $K$, all relations would be at several $\sigma$ from the measurements. Based on this, we can confirm that SU~Cyg must pulsate in its fundamental mode, the pulsation mode cannot be the reason of the discrepancy. We do not have any other explanation about why the P-L relations not calibrated from the Gaia measurements are systematically offset.

\subsection{The Cepheid mass and its companions}
\label{subsection__the_cepheid_mass_and_its_companions}

We measured a very precise dynamical mass for the Cepheid and the companion pair, respectively at \accuracyMcep\,\% and \accuracyMcomp\,\% level. To compare the dynamical mass of the Cepheid with the evolutionary models, we used the average effective temperature $T_\mathrm{eff} = 6165\pm50$\,K and radii $R = 40.6\pm0.8\,R_\odot$ determined by \citet{Trahin_2019_11_0}, who used the SPIPS algorithm to combine all types of available data for a variable star (multi-band and multicolour photometry, radial velocity, effective temperature, and interferometric measurements) in a global modelling of the stellar pulsation. We determined a luminosity $L = 2138\pm109\,L_\odot$, which we compared to \emph{PARSEC} \citep{Bressan_2012_11_0,Nguyen_2022_09_0,Costa_2019_06_0} and \emph{Geneva} stellar evolutionary tracks. Both models follow standard mixing-length theory (MLT) for convective mixing with mixing-length parameter $\alpha_\mathrm{MLT} = 1.6$ and 1.74, respectively for \emph{Geneva} and \emph{Parsec}, and both using the  Schwarzschild criterion to determine convective boundaries. For the convective core overshoot, \emph{Geneva} uses step overshooting with overshoot parameter $\alpha_\mathrm{ov} = 0.1$, while \emph{PARSEC} uses the overshoot formalism of \citet{Bressan_1981_09_0} which would correspond to a step overshooting parameter of 0.25.

\begin{figure*}[!ht]
	\centering
	\resizebox{0.9\hsize}{!}{\includegraphics{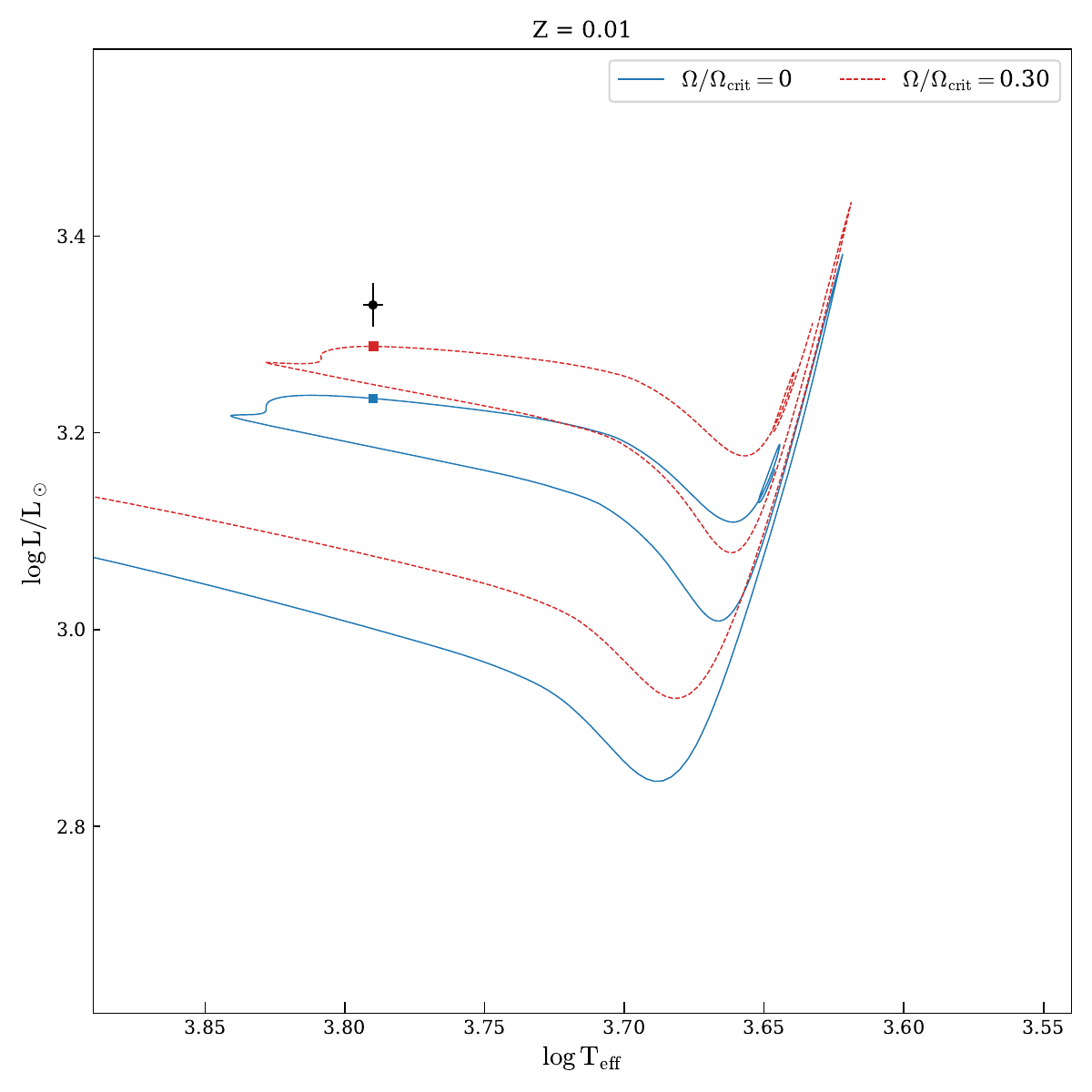}\includegraphics{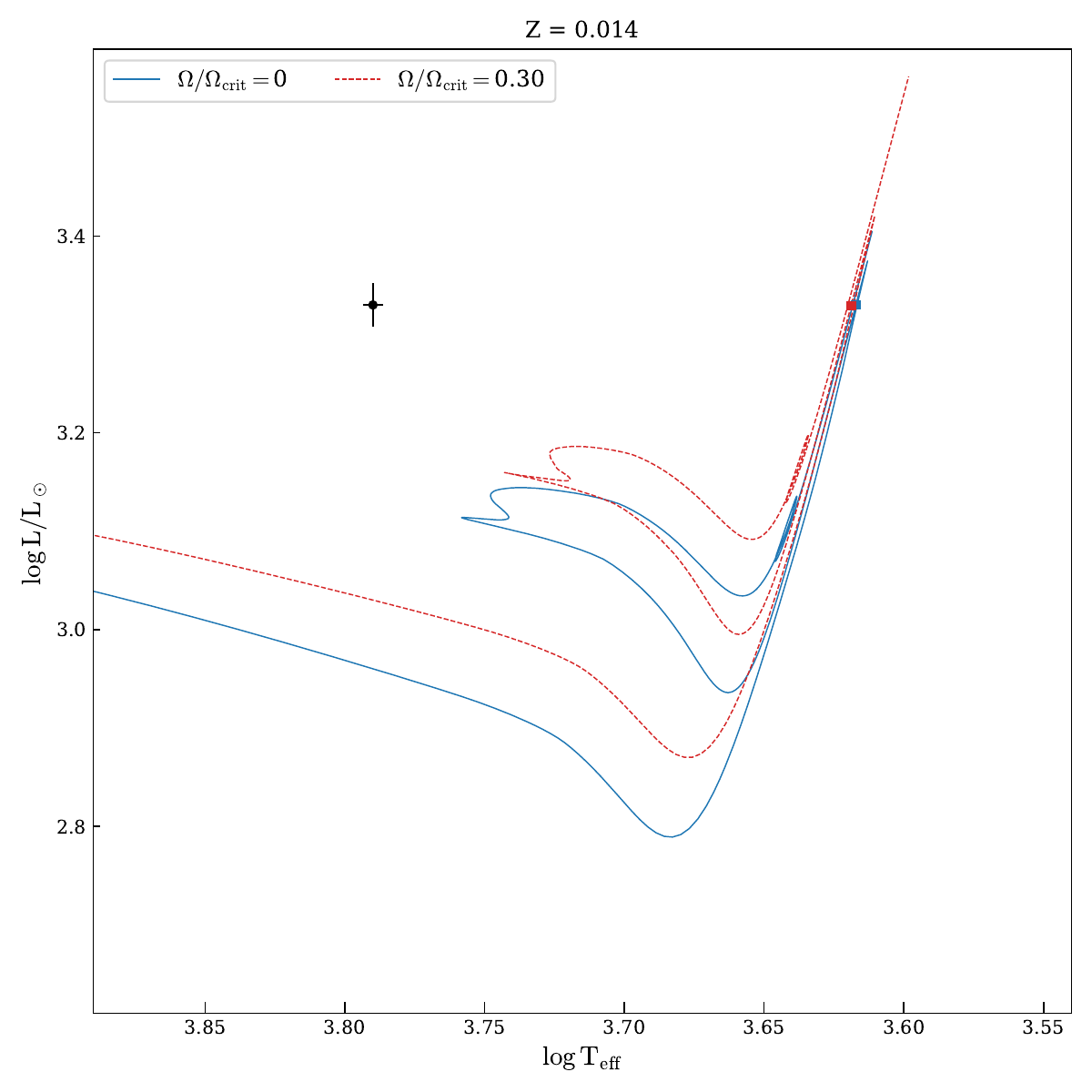}\includegraphics{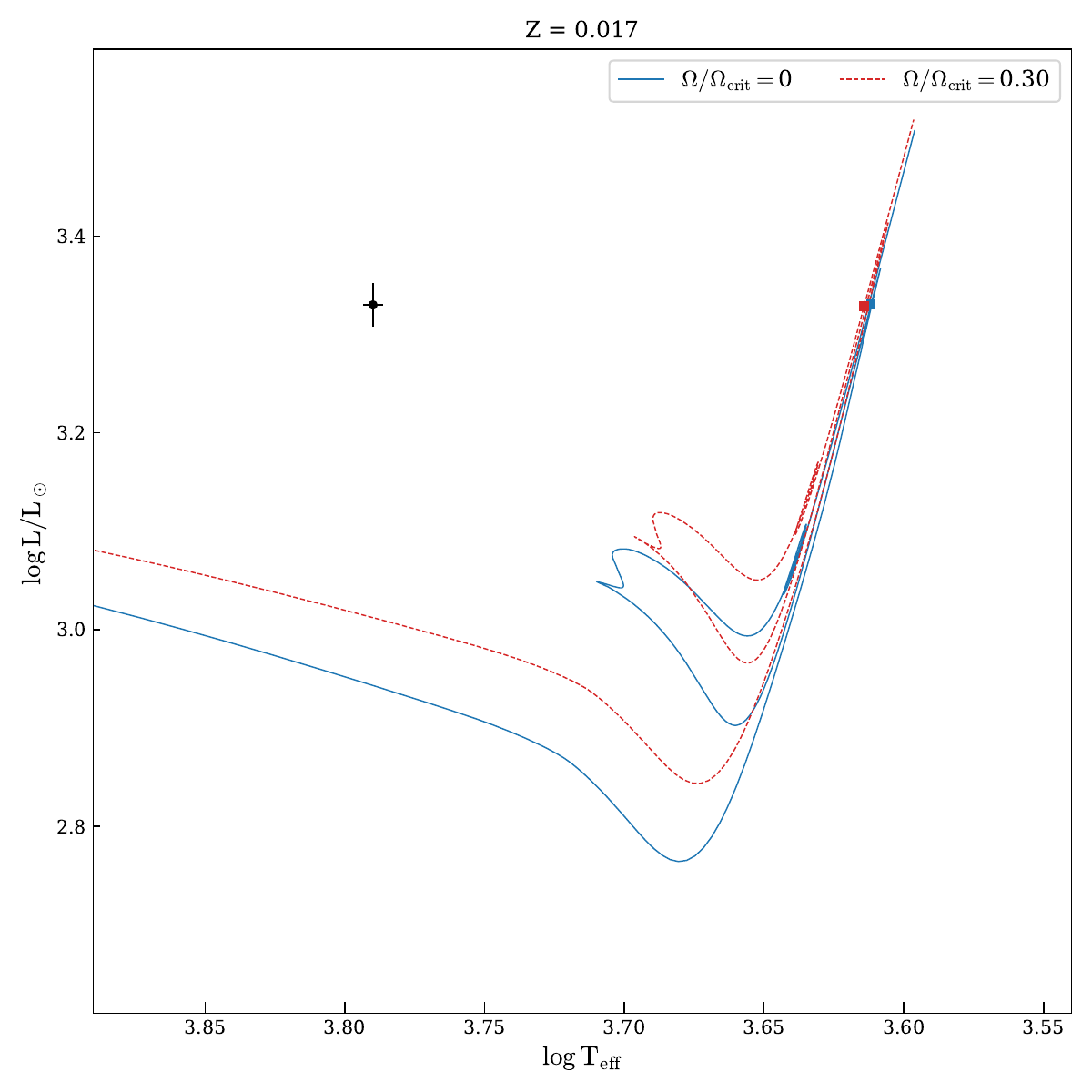}}
	\resizebox{0.9\hsize}{!}{\includegraphics{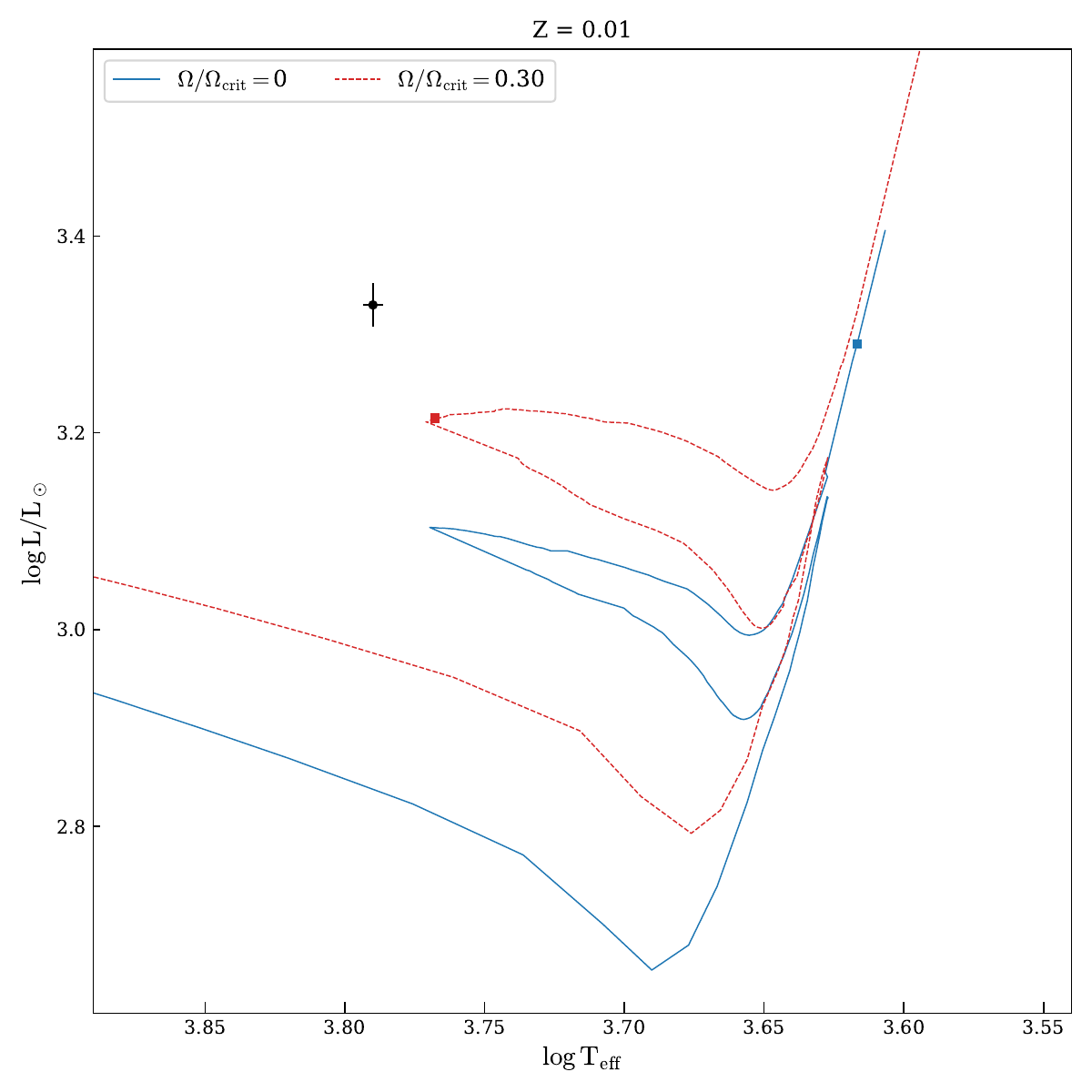}\includegraphics{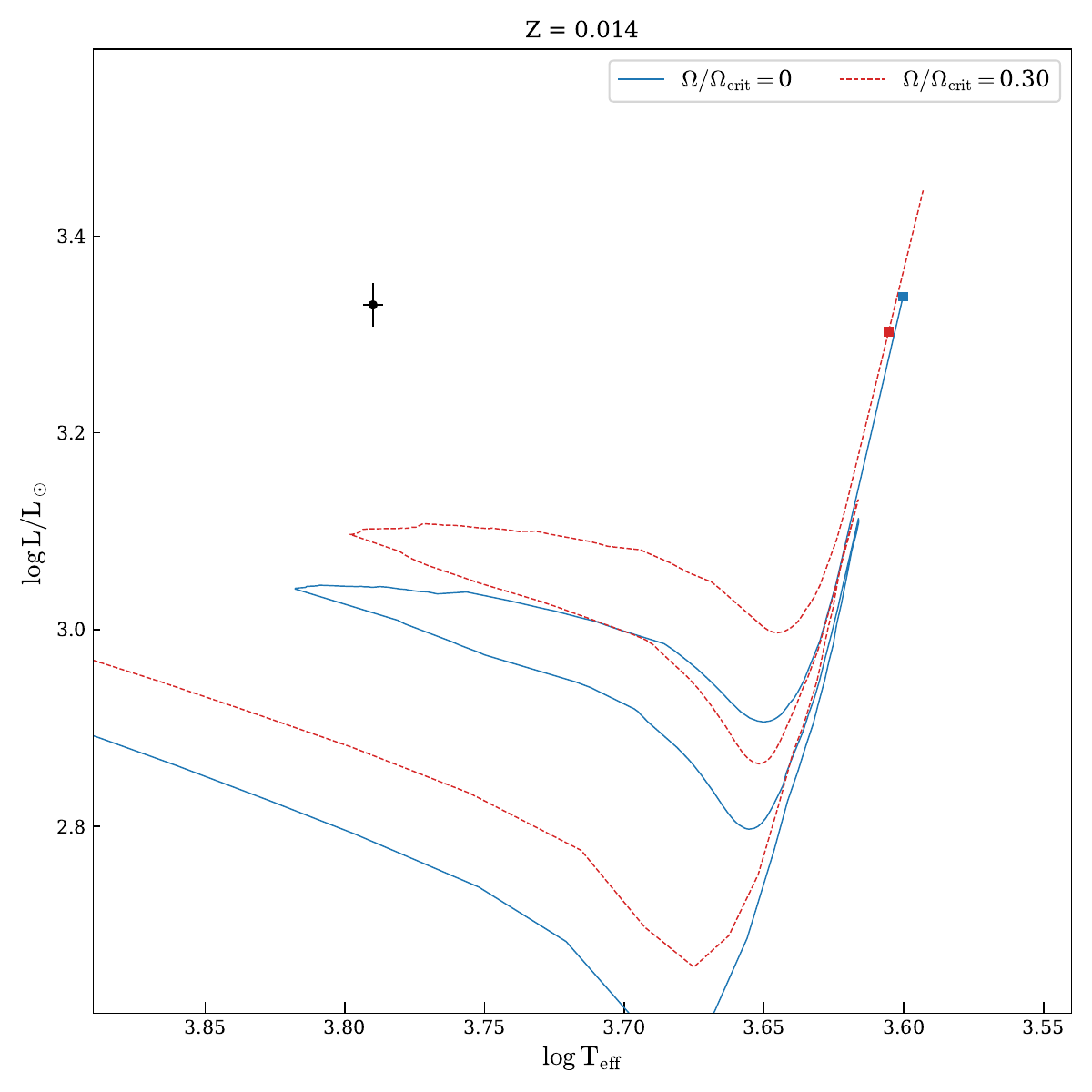}\includegraphics{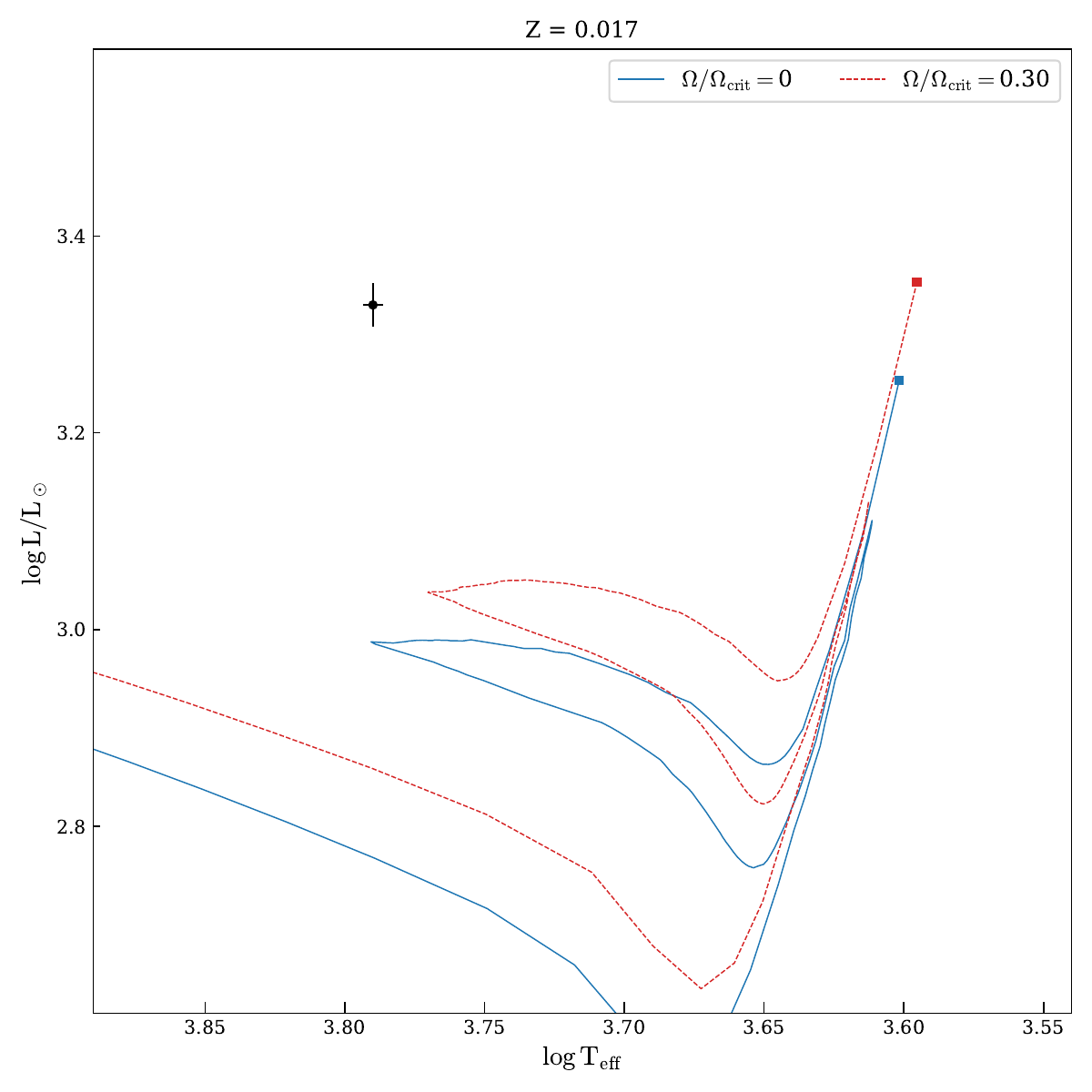}}
	\caption{Comparison of \emph{PARSEC} (top three panels) and \emph{Geneva} (bottom three panels) stellar evolutionary tracks for three metallicities encompassing SU~Cyg metallicity, with (dotted lines) and without rotation for a $4.8\,M_\odot$ stellar mass. The black dot represent the measured value, while the red and blue squares denote the predicted positions for the corresponding track.}
	\label{figure_tracks}
\end{figure*}

The metallicity of SU Cyg is $\sim 0$\,dex \citep{Andrievsky_2013_02_0}, which converts to $Z \sim 0.014$ for both models. We also compared models with metallicities encompassing this value, i.e. 0.01 and 0.017. In Fig.~\ref{figure_tracks} we display tracks for a stellar mass of $4.8\,M_\odot$, without rotation and with a moderate rotation ($\Omega/\Omega_\mathrm{crit} = 0.3$), to see any impact from additional mixing due to rotation. We see that at the expected metallicity of 0.014, both models are inconsistent, including or not rotation effects. Models predict a higher mass than expected, as already reported in our previous works \citep[see e.g.][]{Evans_2018_08_1}. \emph{PARSEC} models predict a mass between $5.1\,M_\odot$ and $5.6\,M_\odot$, while \emph{Geneva} tracks predict $5.4-6.0\,M_\odot$, $> 0.3\,M_\odot$ with our measurement. The only agreement would be with the \emph{PARSEC} models, with and without rotation, with a metallicity of $Z = 0.01$ (see upper right panel of Fig.~\ref{figure_tracks}). However, this would mean $Fe/H \sim -0.18\,$dex which is in a large disagreement with the measured value.

To check if the discrepancy could come from our measured mass, we added 5\,\% error to our astrometric measurements and normalized all RVs to 1\,km~s$^{-1}$. The MCMC analysis gave a similar Cepheid mass with $4.83\pm0.17\,M_\odot$, although a larger uncertainty, making us confident about the robustness of our measurement.

\begin{figure}[!ht]
	\centering
	\resizebox{\hsize}{!}{\includegraphics{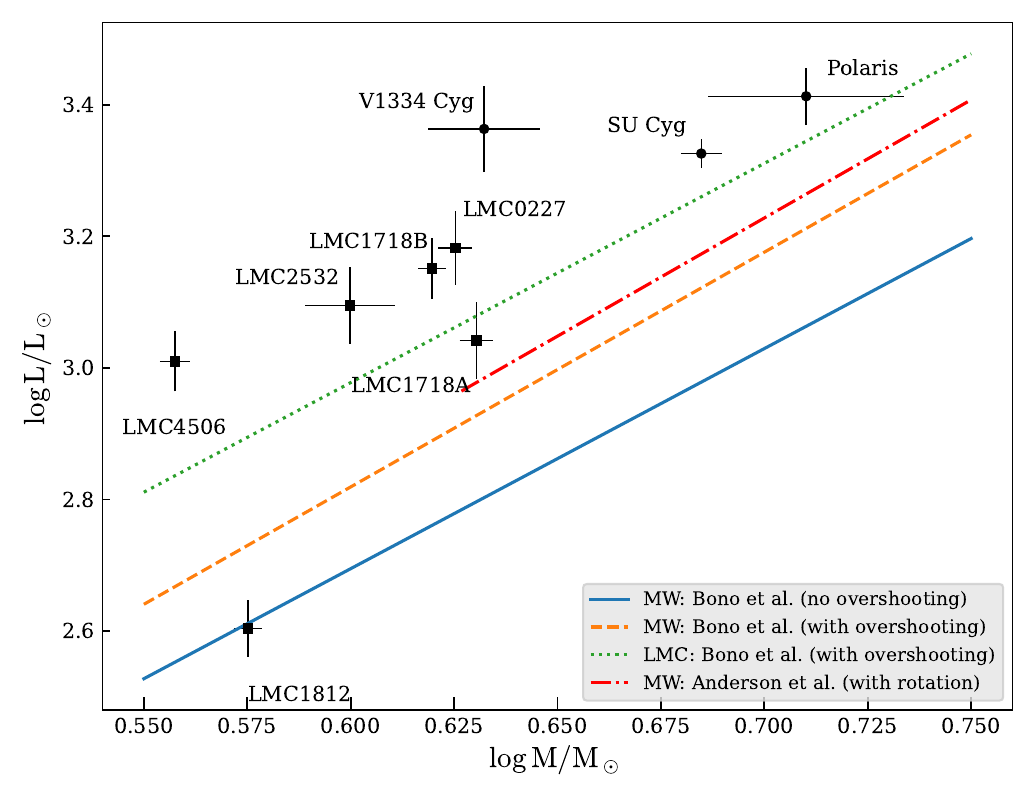}}
	\caption{Cepheid mass-luminosity relation. The MW Cepheids are plotted as filled circles while LMC Cepheids are plotted as squares. Overplotted are predictions from evolutionary tracks.}
	\label{figure_ml_relation}
\end{figure}

The mass of SU~Cyg can be compared with other well-determined masses for the Milky Way (MW) Cepheids, V1334~Cyg \citep{Gallenne_2018_11_0} and \object{Polaris} \citep{Evans_2024_08_0}, and also Cepheids in the Large Magellanic Cloud (LMC) \citep{Pilecki_2018_07_0}. Fig.~\ref{figure_ml_relation} shows them in comparison with predictions from evolutionary tracks from \citet{Bono_2016__0} and \citet{Anderson_2014_04_0}. The Bono tracks cover MW and LMC metallicities (for the MW we used the relations with no main sequence convective core overshoot and with moderate overshoot). The main sequence rotation and moderate convective overshoot are shown for the Anderson relation {\bf(Equ.~5 with solar metallicity)}. The MW Cepheids are consistently brighter than the predictions, as are the LMC Cepheids except for one on the first instability crossing (LMC~1812), and a system which has possibly had binary interaction (LMC~1718~A and B).

The mass for each companion can also be determined if we do the reasonable hypothesis that their orbital inclination is identical to their orbit around the Cepheid. We previously determined the mass function $f_\mathrm{B}$ of the companion pair (see Table~\ref{table__fitted_short_and_long_RV}), which combined with the inclination and the total mass $M_\mathrm{B}$ gives:
\begin{align*}
	M_\mathrm{Bb} &= \dfrac{\sqrt[3]{f_\mathrm{B} M_\mathrm{B}^2}}{\sin i}\\
	M_\mathrm{Ba} &= M_\mathrm{B} - M_\mathrm{Bb}
\end{align*}

We determined a mass for the secondary component of $M_\mathrm{Ba} = \Mb\pm\eMb\,M_\odot$ and for the tertiary star $M_\mathrm{Bb} = \Mc\pm\eMc\,M_\odot$. This would correspond to stars with spectral types of a B8V and F1V star, respectively. This is in good agreement with \citet{Evans_1990_06_0} who suggested a B7.5V star from ultraviolet spectra for the secondary companion, and set an upper limit for the spectral type of the tertiary companion to be later than A0V.

\subsection{The temperature of SU Cyg Ba}

The temperature  of the hottest star in the SU Cyg system can be determined by comparing the IUE spectrum SWP 14773 with the BOSZ Synthetic Stellar Spectral Library \citep{Bohlin_2017_05_0}.  This spectrum has been discussed before \citep{Evans_1995_05_0}, however the model comparisons provide a direct link between temperatures and the masses of detached eclipsing binaries \citep{Evans_2023_09_0}.  Details of the comparisons are provided in \citet{Evans_2024_11_0}.

The first step in the temperature determination is obtaining the reddening $E(B-V)$. Since this was discussed by \citep{Evans_1995_05_0} using colours corrected for the companion, we start with the $E(B-V) = 0.109$ from \citet{Trahin_2019_11_0}. Using this, we dereddened the IUE spectrum and ran it thought the program which compares it with models of a series of temperatures. Details of the fitting are provided by the figures in Appendix~\ref{appendix__Comparison_of_SU_Cyg_B_spectrum_with_models}.

The best fit temperature is $13500 \pm 920$\,K. from the parabola fit to the standard deviations of the spectrum-model differences (Fig.~\ref{figure_sm}). As in the discussion of the DEBs, the uncertainty was also estimated visually from the spectral differences (Fig.~\ref{figure_sm_diff_single}) to be 500\,K. This corresponds approximately to a B7.3V in the calibration of \citet{Pecaut_2013_09_0}, which is in good agreement with our estimate from the previous Section. To estimate its mass, we can use the formula relating temperature and mass derived from detached eclipsing binaries (DEBs) by \citet{Evans_2023_09_0}. As for the \object{FN~Vel} system \citep{Evans_2024_11_0}, because SU~Cyg~Ba is the companion of a massive young Cepheid, it is at the younger (hotter) half of the relation. Thus the mass is decreased by 0.02 in $\log{M}$. SU~Cyg~Ba is a HgMn star and in fact the DEBs study provides some information about those. In the sample of DEBs there are 5 Am (metallic lined) stars. They sit above (at larger masses) than normal stars. Although they are all cooler than SU Cyg Ba, this is an indication that chemically peculiar stars have cooler temperatures than normal stars. We can make an approximate correction of 0.02 in $\log{T}$ for this, the mass of Ba becomes $3.75\,M_\odot$, and the mass of Bb is $1.37\,M_\odot$. They are similar to our previous determination of Sect.~\ref{subsection__the_cepheid_mass_and_its_companions}.

\section{Conclusion}
\label{section__conclusion}
 
 We report new interferometric and ultraviolet spectroscopic observation of the binary Cepheid SU Cygni. These measurements enabled us to detect the astrometric motion of the centre of light of the binary companion and to measure their radial velocities. Combining with RVs from the literature for the Cepheid, we were able to measure the mass of the Cepheid to \accuracyMcep\,\% and its distance to \accuracydist\,\%. Prediction from evolutionary models do not agree with the current luminosity and mass of the Cepheid, providing larger values than expected. We already report such disagreement in our previous works for the Cepheid V1334 Cyg and Polaris. Accounting for rotation or core overshooting do not solve this issue.
 
 This work also provide the most precise distance for a Cepheid. We found a slight disagreement with Gaia at $1.5\sigma$, which is acceptable as SU Cyg is very bright and likely saturate the camera. We used our measurement to estimate the absolute magnitude of the star which we compared with prediction from a few P-L relations, calibrated from photometry or parallax measurements. We found that the photometry-calibrated P-L relation are in disagreement with our measurement, while the parallax-calibrated relations are in better agreement. 
 
 SU Cyg is our best Cepheid in our sample as it provides the most precise mass and distance estimate. With our similar previous work on V1334 Cyg and Polaris, we now have a few Cepheid with precise dynamical mass measurement which may be used to calibrate stellar evolution models. Finally, V1334 Cyg and SU Cyg are bright targets and pulsating stars, therefore they can serve as check stars for the bright-star mode of Gaia and test the saturation effect corrections, in addition to checking for systematics from colour variation effects.

 
 \begin{acknowledgements}
 	This work is based upon observations obtained with the Georgia State University Center for High Angular Resolution Astronomy Array at Mount Wilson Observatory. The CHARA Array is supported by the National Science Foundation under Grant No. AST-1636624 and AST-2034336. Institutional support has been provided from the GSU College of Arts and Sciences and the GSU Office of the Vice President for Research and Economic Development. Time at the CHARA Array was granted through the NOIRLab community access program (NOAO ProgID 2019A-N0071-PI and NOIRLab PropID 2023B-381062; PI: A. Gallenne). AG acknowledges the support of the Agencia Nacional de Investigaci\'on Cient\'ifica y Desarrollo (ANID) through the FONDECYT Regular grant 1241073 and ANID-ALMA fund No. ASTRO20-0059. PK, GP and WG acknowledge funding from the European Research Council (ERC) under the European Union’s Horizon 2020 research and innovation program (project UniverScale, grant agreement 951549). JDM acknowledges funding for the development of MIRC-X (NASA-XRP NNX16AD43G, NSF-AST 1909165). Support was provided by HST-GO-16208.001-A for CP and JK. Support for EMW was provided from HST-GO-14194.001-A. Support was provided by HST-GO-13841 for JDM and CP. SK acknowledges funding for MIRC-X received funding from the European Research Council (ERC) under the European Union's Horizon 2020 research and innovation programme (Starting Grant No. 639889 and Consolidated Grant No. 101003096). GP acknowledges financial support from  the Polish Ministry of Science and Higher Education with the grant agreement 2024/WK/02. GP and PK acknowledge support from the Polish-French Marie Skłodowska-Curie and Pierre Curie Science Prize awarded by the Foundation for Polish Science. B.P. acknowledges funding from the Polish National Science Center grant SONATA BIS 2020/38/E/ST9/00486. 
 	This work has made use of data from the European Space Agency (ESA) mission {\it Gaia} (\url{https://www.cosmos.esa.int/gaia}), processed by the {\it Gaia} Data Processing and Analysis Consortium (DPAC, \url{https://www.cosmos.esa.int/web/gaia/dpac/consortium}). Funding for the DPAC has been provided by national institutions, in particular the institutions participating in the {\it Gaia} Multilateral Agreement. This research made use of services provided by the Jean-Marie Mariotti Center (Aspro and SearchCal). The SIMBAD database, and NASA's Astrophysics Data System Bibliographic Services were used in the preparation of this paper.
 \end{acknowledgements}

 
 \bibliographystyle{aa}   
\bibliography{bibliographie}

\begin{thebibliography}{87}
\expandafter\ifx\csname natexlab\endcsname\relax\def\natexlab#1{#1}\fi

\bibitem[{{Alcock} {et~al.}(1995){Alcock}, {Allsman}, {Axelrod}, {Bennett},
  {Cook}, {Freeman}, {Griest}, {Marshall}, {Peterson}, {Pratt}, {Quinn},
  {Reimann}, {Rodgers}, {Stubbs}, {Sutherland}, \& {Welch}}]{Alcock_1995_04_0}
{Alcock}, C., {Allsman}, R.~A., {Axelrod}, T.~S., {et~al.} 1995, \aj, 109, 1653

\bibitem[{{Anderson} {et~al.}(2014){Anderson}, {Ekstr{\"o}m}, {Georgy},
  {Meynet}, {Mowlavi}, \& {Eyer}}]{Anderson_2014_04_0}
{Anderson}, R.~I., {Ekstr{\"o}m}, S., {Georgy}, C., {et~al.} 2014, \aap, 564,
  A100

\bibitem[{{Andrievsky} {et~al.}(2013){Andrievsky}, {L{\'e}pine}, {Korotin},
  {Luck}, {Kovtyukh}, \& {Maciel}}]{Andrievsky_2013_02_0}
{Andrievsky}, S.~M., {L{\'e}pine}, J.~R.~D., {Korotin}, S.~A., {et~al.} 2013,
  \mnras, 428, 3252

\bibitem[{{Anugu} {et~al.}(2020){Anugu}, {Le Bouquin}, {Monnier}, {Kraus},
  {Setterholm}, {Labdon}, {Davies}, {Lanthermann}, {Gardner}, {Ennis},
  {Johnson}, {Ten Brummelaar}, {Schaefer}, \& {Sturmann}}]{Anugu_2020_10_0}
{Anugu}, N., {Le Bouquin}, J.-B., {Monnier}, J.~D., {et~al.} 2020, \aj, 160,
  158

\bibitem[{{Baade}(1926)}]{Baade_1926_11_0}
{Baade}, W. 1926, Astronomische Nachrichten, 228, 359

\bibitem[{{Baer} {et~al.}(2018){Baer}, {Proffitt}, \&
  {Lockwood}}]{Baer_2018_01_0}
{Baer}, M., {Proffitt}, C.~R., \& {Lockwood}, S.~A. 2018, {A Python Script for
  Aligning the STIS Echelle Blaze Function}, Instrument Science Report STIS
  2018-1, 13 pages

\bibitem[{{Barnes} {et~al.}(1987){Barnes}, {Moffett}, \&
  {Slovak}}]{Barnes_1987_10_1}
{Barnes}, Thomas~G., I., {Moffett}, T.~J., \& {Slovak}, M.~H. 1987, \apjs, 65,
  307

\bibitem[{{Benedict} {et~al.}(2007){Benedict}, {McArthur}, {Feast}, {Barnes},
  {Harrison}, {Patterson}, {Menzies}, {Bean}, \&
  {Freedman}}]{Benedict_2007_04_0}
{Benedict}, G.~F., {McArthur}, B.~E., {Feast}, M.~W., {et~al.} 2007, \aj, 133,
  1810

\bibitem[{{Berdnikov}(2008)}]{Berdnikov_2008_04_0}
{Berdnikov}, L.~N. 2008, VizieR Online Data Catalog: II/285, originally
  published in: Sternberg Astronomical Institute, Moscow, 2285

\bibitem[{{Boffin} {et~al.}(2014){Boffin}, {Hillen}, {Berger}, {Jorissen},
  {Blind}, {Le Bouquin}, {Miko{\l}ajewska}, \& {Lazareff}}]{Boffin_2014_04_0}
{Boffin}, H.~M.~J., {Hillen}, M., {Berger}, J.~P., {et~al.} 2014, \aap, 564, A1

\bibitem[{{Bohlin} {et~al.}(2017){Bohlin}, {M{\'e}sz{\'a}ros}, {Fleming},
  {Gordon}, {Koekemoer}, \& {Kov{\'a}cs}}]{Bohlin_2017_05_0}
{Bohlin}, R.~C., {M{\'e}sz{\'a}ros}, S., {Fleming}, S.~W., {et~al.} 2017, \aj,
  153, 234

\bibitem[{{Bono} {et~al.}(2016){Bono}, {Braga}, {Pietrinferni}, {Magurno},
  {Dall'Ora}, {Fiorentino}, {Fukue}, {Inno}, {Marengo}, {Bergemann},
  {Buonanno}, {da Silva}, {Fabrizio}, {Ferraro}, {Gilmozzi}, {Iannicola},
  {Kausch}, {Kobayashi}, {Kovtyukh}, {Lemasle}, {Marconi}, {Marinoni},
  {Marrese}, {Mart{\'{\i}}nez-V{\'a}zquez}, {Matsunaga}, {Monelli}, {Neeley},
  {Nonino}, {Proxauf}, {Przybilla}, {Romaniello}, {Salaris}, {Sneden},
  {Stetson}, {Th{\'e}venin}, {Tsujimoto}, {Urbaneja}, {Valenti}, \&
  {Zoccali}}]{Bono_2016__0}
{Bono}, G., {Braga}, V.~F., {Pietrinferni}, A., {et~al.} 2016, \memsai, 87, 358

\bibitem[{{Bono} {et~al.}(2006){Bono}, {Caputo}, \&
  {Castellani}}]{Bono_2006__0}
{Bono}, G., {Caputo}, F., \& {Castellani}, V. 2006, \memsai, 77, 207

\bibitem[{{Borgniet} {et~al.}(2019){Borgniet}, {Kervella}, {Nardetto},
  {Gallenne}, {M{\'e}rand}, {Anderson}, {Aufdenberg}, {Breuval}, {Gieren},
  {Hocd{\'e}}, {Javanmardi}, {Lagadec}, {Pietrzy{\'n}ski}, \&
  {Trahin}}]{Borgniet_2019_11_0}
{Borgniet}, S., {Kervella}, P., {Nardetto}, N., {et~al.} 2019, \aap, 631, A37

\bibitem[{{Bressan} {et~al.}(2012){Bressan}, {Marigo}, {Girardi}, {Salasnich},
  {Dal Cero}, {Rubele}, \& {Nanni}}]{Bressan_2012_11_0}
{Bressan}, A., {Marigo}, P., {Girardi}, L., {et~al.} 2012, \mnras, 427, 127

\bibitem[{{Bressan} {et~al.}(1981){Bressan}, {Chiosi}, \&
  {Bertelli}}]{Bressan_1981_09_0}
{Bressan}, A.~G., {Chiosi}, C., \& {Bertelli}, G. 1981, \aap, 102, 25

\bibitem[{{Breuval} {et~al.}(2020){Breuval}, {Kervella}, {Anderson}, {Riess},
  {Arenou}, {Trahin}, {M{\'e}rand}, {Gallenne}, {Gieren}, {Storm}, {Bono},
  {Pietrzy{\'n}ski}, {Nardetto}, {Javanmardi}, \&
  {Hocd{\'e}}}]{Breuval_2020_11_0}
{Breuval}, L., {Kervella}, P., {Anderson}, R.~I., {et~al.} 2020, \aap, 643,
  A115

\bibitem[{{Chelli} {et~al.}(2016){Chelli}, {Duvert}, {Bourg{\`e}s}, {Mella},
  {Lafrasse}, {Bonneau}, \& {Chesneau}}]{Chelli_2016_05_0}
{Chelli}, A., {Duvert}, G., {Bourg{\`e}s}, L., {et~al.} 2016, \aap, 589, A112

\bibitem[{{Clementini} {et~al.}(2019){Clementini}, {Ripepi}, {Molinaro},
  {Garofalo}, {Muraveva}, {Rimoldini}, {Guy}, {Jevardat de Fombelle},
  {Nienartowicz}, {Marchal}, {Audard}, {Holl}, {Leccia}, {Marconi}, {Musella},
  {Mowlavi}, {Lecoeur-Taibi}, {Eyer}, {De Ridder}, {Regibo}, {Sarro},
  {Szabados}, {Evans}, \& {Riello}}]{Clementini_2019_02_0}
{Clementini}, G., {Ripepi}, V., {Molinaro}, R., {et~al.} 2019, \aap, 622, A60

\bibitem[{{Costa} {et~al.}(2019){Costa}, {Girardi}, {Bressan}, {Marigo},
  {Rodrigues}, {Chen}, {Lanza}, \& {Goudfrooij}}]{Costa_2019_06_0}
{Costa}, G., {Girardi}, L., {Bressan}, A., {et~al.} 2019, \mnras, 485, 4641

\bibitem[{{Czesla} {et~al.}(2019){Czesla}, {Schr{\"o}ter}, {Schneider},
  {Huber}, {Pfeifer}, {Andreasen}, \& {Zechmeister}}]{Czesla_2019_06_0}
{Czesla}, S., {Schr{\"o}ter}, S., {Schneider}, C.~P., {et~al.} 2019, {PyA:
  Python astronomy-related packages}, Astrophysics Source Code Library, record
  ascl:1906.010

\bibitem[{{Evans}(1988)}]{Evans_1988_03_0}
{Evans}, N.~R. 1988, \apjs, 66, 343

\bibitem[{{Evans}(1989)}]{Evans_1989_06_0}
{Evans}, N.~R. 1989, \aj, 97, 1737

\bibitem[{{Evans}(1995)}]{Evans_1995_05_0}
{Evans}, N.~R. 1995, \apj, 445, 393

\bibitem[{{Evans} \& {Bolton}(1990)}]{Evans_1990_06_0}
{Evans}, N.~R. \& {Bolton}, C.~T. 1990, \apj, 356, 630

\bibitem[{{Evans} {et~al.}(2023){Evans}, {Ferrari}, {Kuraszkiewicz},
  {Silverberg}, {Nichols}, {Torres}, \& {Fischbach}}]{Evans_2023_09_0}
{Evans}, N.~R., {Ferrari}, M.~G., {Kuraszkiewicz}, J., {et~al.} 2023, \aj, 166,
  109

\bibitem[{{Evans} {et~al.}(2018{\natexlab{a}}){Evans}, {Karovska}, {Bond},
  {Schaefer}, {Sahu}, {Mack}, {Nelan}, {Gallenne}, \&
  {Tingle}}]{Evans_2018_08_1}
{Evans}, N.~R., {Karovska}, M., {Bond}, H.~E., {et~al.} 2018{\natexlab{a}},
  \apj, 863, 187

\bibitem[{{Evans} {et~al.}(2024{\natexlab{a}}){Evans}, {Kervella},
  {Kuraszkiewicz}, {G{\"u}nther}, {Anderson}, {Proffitt}, {Gallenne},
  {M{\'e}rand}, {Trahin}, {Viviani}, \& {Shetye}}]{Evans_2024_11_0}
{Evans}, N.~R., {Kervella}, P., {Kuraszkiewicz}, J., {et~al.}
  2024{\natexlab{a}}, \aj, 168, 221

\bibitem[{{Evans} {et~al.}(2018{\natexlab{b}}){Evans}, {Proffitt}, {Carpenter},
  {Winston}, {Kober}, {G{\"u}nther}, {Gorynya}, {Rastorguev}, \&
  {Inno}}]{Evans_2018_10_0}
{Evans}, N.~R., {Proffitt}, C., {Carpenter}, K.~G., {et~al.}
  2018{\natexlab{b}}, \apj, 866, 30

\bibitem[{{Evans} {et~al.}(2024{\natexlab{b}}){Evans}, {Schaefer}, {Gallenne},
  {Torres}, {Horch}, {Anderson}, {Monnier}, {Roettenbacher}, {Baron}, {Anugu},
  {Davidson}, {Kervella}, {Bras}, {Proffitt}, {M{\'e}rand}, {Karovska},
  {Jones}, {Lanthermann}, {Kraus}, {Codron}, {Bond}, \&
  {Viviani}}]{Evans_2024_08_0}
{Evans}, N.~R., {Schaefer}, G.~H., {Gallenne}, A., {et~al.} 2024{\natexlab{b}},
  \apj, 971, 190

\bibitem[{{Feast} \& {Catchpole}(1997)}]{Feast_1997_03_0}
{Feast}, M.~W. \& {Catchpole}, R.~M. 1997, \mnras, 286, L1

\bibitem[{{Fernie}(1990)}]{Fernie_1990_01_0}
{Fernie}, J.~D. 1990, \apjs, 72, 153

\bibitem[{{Foreman-Mackey} {et~al.}(2013){Foreman-Mackey}, {Hogg}, {Lang}, \&
  {Goodman}}]{Foreman-Mackey_2013_03_0}
{Foreman-Mackey}, D., {Hogg}, D.~W., {Lang}, D., \& {Goodman}, J. 2013, \pasp,
  125, 306

\bibitem[{{Fouqu{\'e}} {et~al.}(2007){Fouqu{\'e}}, {Arriagada}, {Storm},
  {Barnes}, {Nardetto}, {M{\'e}rand}, {Kervella}, {Gieren}, {Bersier},
  {Benedict}, \& {McArthur}}]{Fouque_2007_12_0}
{Fouqu{\'e}}, P., {Arriagada}, P., {Storm}, J., {et~al.} 2007, \aap, 476, 73

\bibitem[{{Freedman}(2021)}]{Freedman_2021_09_3}
{Freedman}, W.~L. 2021, \apj, 919, 16

\bibitem[{{Gallenne} {et~al.}(2018){Gallenne}, {Kervella}, {Evans}, {Proffitt},
  {Monnier}, {M{\'e}rand}, {Nelan}, {Winston}, {Pietrzy{\'n}ski}, {Schaefer},
  {Gieren}, {Anderson}, {Borgniet}, {Kraus}, {Roettenbacher}, {Baron},
  {Pilecki}, {Taormina}, {Graczyk}, {Mowlavi}, \& {Eyer}}]{Gallenne_2018_11_0}
{Gallenne}, A., {Kervella}, P., {Evans}, N.~R., {et~al.} 2018, \apj, 867, 121

\bibitem[{{Gallenne} {et~al.}(2012){Gallenne}, {Kervella}, \&
  {M{\'e}rand}}]{Gallenne_2012_02_0}
{Gallenne}, A., {Kervella}, P., \& {M{\'e}rand}, A. 2012, \aap, 538, A24

\bibitem[{{Gallenne} {et~al.}(2017){Gallenne}, {Kervella}, {M{\'e}rand},
  {Pietrzy{\'n}ski}, {Gieren}, {Nardetto}, \& {Trahin}}]{Gallenne_2017_11_0}
{Gallenne}, A., {Kervella}, P., {M{\'e}rand}, A., {et~al.} 2017, \aap, 608, A18

\bibitem[{{Gallenne} {et~al.}(2013){Gallenne}, {M{\'e}rand}, {Kervella},
  {Chesneau}, {Breitfelder}, \& {Gieren}}]{Gallenne_2013_10_0}
{Gallenne}, A., {M{\'e}rand}, A., {Kervella}, P., {et~al.} 2013, \aap, 558,
  A140

\bibitem[{{Gallenne} {et~al.}(2015){Gallenne}, {M{\'e}rand}, {Kervella},
  {Monnier}, {Schaefer}, {Baron}, {Breitfelder}, {Le Bouquin}, {Roettenbacher},
  {Gieren}, {Pietrzy{\'n}ski}, {McAlister}, {ten Brummelaar}, {Sturmann},
  {Sturmann}, {Turner}, {Ridgway}, \& {Kraus}}]{Gallenne_2015_07_0}
{Gallenne}, A., {M{\'e}rand}, A., {Kervella}, P., {et~al.} 2015, \aap, 579, A68

\bibitem[{{Gallenne} {et~al.}(2021){Gallenne}, {M{\'e}rand}, {Kervella},
  {Pietrzy{\'n}ski}, {Gieren}, {Hocd{\'e}}, {Breuval}, {Nardetto}, \&
  {Lagadec}}]{Gallenne_2021_07_3}
{Gallenne}, A., {M{\'e}rand}, A., {Kervella}, P., {et~al.} 2021, \aap, 651,
  A113

\bibitem[{{Groenewegen}(2013)}]{Groenewegen_2013_02_0}
{Groenewegen}, M.~A.~T. 2013, \aap, 550, A70

\bibitem[{{Groenewegen}(2018)}]{Groenewegen_2018_11_0}
{Groenewegen}, M.~A.~T. 2018, \aap, 619, A8

\bibitem[{{Halbwachs} {et~al.}(2023){Halbwachs}, {Pourbaix}, {Arenou},
  {Galluccio}, {Guillout}, {Bauchet}, {Marchal}, {Sadowski}, \&
  {Teyssier}}]{Halbwachs_2023_06_0}
{Halbwachs}, J.-L., {Pourbaix}, D., {Arenou}, F., {et~al.} 2023, \aap, 674, A9

\bibitem[{{Hellerich}(1919)}]{Hellerich_1919_11_0}
{Hellerich}, J. 1919, Astronomische Nachrichten, 210, 65

\bibitem[{{Hocd{\'e}} {et~al.}(2020{\natexlab{a}}){Hocd{\'e}}, {Nardetto},
  {Borgniet}, {Lagadec}, {Kervella}, {M{\'e}rand}, {Evans}, {Gillet},
  {Mathias}, {Chiavassa}, {Gallenne}, {Breuval}, \&
  {Javanmardi}}]{Hocde_2020_09_0}
{Hocd{\'e}}, V., {Nardetto}, N., {Borgniet}, S., {et~al.} 2020{\natexlab{a}},
  \aap, 641, A74

\bibitem[{{Hocd{\'e}} {et~al.}(2020{\natexlab{b}}){Hocd{\'e}}, {Nardetto},
  {Lagadec}, {Niccolini}, {Domiciano de Souza}, {M{\'e}rand}, {Kervella},
  {Gallenne}, {Marengo}, {Trahin}, {Gieren}, {Pietrzy{\'n}ski}, {Borgniet},
  {Breuval}, \& {Javanmardi}}]{Hocde_2020_01_0}
{Hocd{\'e}}, V., {Nardetto}, N., {Lagadec}, E., {et~al.} 2020{\natexlab{b}},
  \aap, 633, A47

\bibitem[{{Hocd{\'e}} {et~al.}(2021){Hocd{\'e}}, {Nardetto}, {Matter},
  {Lagadec}, {M{\'e}rand}, {Cruzal{\`e}bes}, {Meilland}, {Millour}, {Lopez},
  {Berio}, {Weigelt}, {Petrov}, {Isbell}, {Jaffe}, {Kervella}, {Glindemann},
  {Sch{\"o}ller}, {Allouche}, {Gallenne}, {Domiciano de Souza}, {Niccolini},
  {Kokoulina}, {Varga}, {Lagarde}, {Augereau}, {van Boekel}, {Bristow},
  {Henning}, {Hofmann}, {Zins}, {Danchi}, {Delbo}, {Dominik}, {G{\'a}mez
  Rosas}, {Klarmann}, {Hron}, {Hogerheijde}, {Meisenheimer}, {Pantin},
  {Paladini}, {Robbe-Dubois}, {Schertl}, {Stee}, {Waters}, {Lehmitz},
  {Bettonvil}, {Heininger}, {Bristow}, {Woillez}, {Wolf}, {Yoffe}, {Szabados},
  {Chiavassa}, {Borgniet}, {Breuval}, {Javanmardi}, {{\'A}brah{\'a}m},
  {Abadie}, {Abuter}, {Accardo}, {Adler}, {Ag{\'o}cs}, {Alonso}, {Antonelli},
  {B{\"o}hm}, {Bailet}, {Bazin}, {Beckmann}, {Beltran}, {Boland}, {Bourget},
  {Brast}, {Bresson}, {Burtscher}, {Buter}, {Castillo}, {Chelli}, {Cid},
  {Clausse}, {Connot}, {Conzelmann}, {De Haan}, {Ebert}, {Elswijk}, {Fantei},
  {Frahm}, {G{\'a}mez Rosas}, {Gabasch}, {Garces}, {Girard}, {Glazenborg},
  {Gont{\'e}}, {Gonz{\'a}lez Herrera}, {Graser}, {Guajardo}, {Guitton},
  {Hanenburg}, {Haubois}, {Hubin}, {Huerta}, {Idserda}, {Ives}, {Jakob},
  {Jask{\'o}}, {Jochum}, {Klein}, {Kragt}, {Kroes}, {Kuindersma}, {Labadie},
  {Laun}, {Le Poole}, {Leinert}, {Lizon}, {Lopez}, {Marcotto}, {Mauclert},
  {Maurer}, {Mehrgan}, {Meisner}, {Meixner}, {Mellein}, {Mohr}, {Morel},
  {Mosoni}, {Navarro}, {Neumann}, {Nu{\ss}baum}, {Pallanca}, {Pasquini},
  {Percheron}, {Phan Duc}, {Pott}, {Pozna}, {Ridinger}, {Rigal}, {Riquelme},
  {Rivinius}, {Roelfsema}, {Rohloff}, {Rousseau}, {Schuhler}, {Schuil},
  {Shabun}, {Soulain}, {Stephan}, {ter Horst}, {Tromp}, {Vakili}, {van Duin},
  {Venema}, {Vinther}, {Wittkowski}, \& {Wrhel}}]{Hocde_2021_07_9}
{Hocd{\'e}}, V., {Nardetto}, N., {Matter}, A., {et~al.} 2021, \aap, 651, A92

\bibitem[{{Imbert}(1984)}]{Imbert_1984_12_0}
{Imbert}, M. 1984, \aaps, 58, 529

\bibitem[{{Irwin}(1952)}]{Irwin_1952_07_0}
{Irwin}, J.~B. 1952, \apj, 116, 211

\bibitem[{{Irwin}(1959)}]{Irwin_1959_05_0}
{Irwin}, J.~B. 1959, \aj, 64, 149

\bibitem[{{Keller}(2008)}]{Keller_2008_04_0}
{Keller}, S.~C. 2008, \apj, 677, 483

\bibitem[{{Kervella} {et~al.}(2006){Kervella}, {M{\'e}rand}, {Perrin}, \&
  {Coud{\'e} Du Foresto}}]{Kervella_2006_03_0}
{Kervella}, P., {M{\'e}rand}, A., {Perrin}, G., \& {Coud{\'e} Du Foresto}, V.
  2006, \aap, 448, 623

\bibitem[{{Kiss}(1998)}]{Kiss_1998_07_0}
{Kiss}, L.~L. 1998, \mnras, 297, 825

\bibitem[{{Konacki} {et~al.}(2010){Konacki}, {Muterspaugh}, {Kulkarni}, \&
  {He{\l}miniak}}]{Konacki_2010_08_0}
{Konacki}, M., {Muterspaugh}, M.~W., {Kulkarni}, S.~R., \& {He{\l}miniak},
  K.~G. 2010, \apj, 719, 1293

\bibitem[{{Kovtyukh} {et~al.}(2016){Kovtyukh}, {Lemasle}, {Chekhonadskikh},
  {Bono}, {Matsunaga}, {Yushchenko}, {Anderson}, {Belik}, {da Silva}, \&
  {Inno}}]{Kovtyukh_2016_08_0}
{Kovtyukh}, V., {Lemasle}, B., {Chekhonadskikh}, F., {et~al.} 2016, \mnras,
  460, 2077

\bibitem[{{Lindegren} {et~al.}(2021){Lindegren}, {Bastian}, {Biermann},
  {Bombrun}, {de Torres}, {Gerlach}, {Geyer}, {Hern{\'a}ndez}, {Hilger},
  {Hobbs}, {Klioner}, {Lammers}, {McMillan}, {Ramos-Lerate},
  {Steidelm{\"u}ller}, {Stephenson}, \& {van Leeuwen}}]{Lindegren_2021_05_7}
{Lindegren}, L., {Bastian}, U., {Biermann}, M., {et~al.} 2021, \aap, 649, A4

\bibitem[{{Luck}(2018)}]{Luck_2018_10_0}
{Luck}, R.~E. 2018, \aj, 156, 171

\bibitem[{{Mayor} {et~al.}(1983){Mayor}, {Imbert}, {Andersen}, {Ardeberg},
  {Baranne}, {Benz}, {Ischi}, {Lindgren}, {Martin}, {Maurice}, {Nordstrom}, \&
  {Prevot}}]{Mayor_1983_12_0}
{Mayor}, M., {Imbert}, M., {Andersen}, J., {et~al.} 1983, \aaps, 54, 495

\bibitem[{{M{\'e}rand} {et~al.}(2015){M{\'e}rand}, {Kervella}, {Breitfelder},
  {Gallenne}, {Coud{\'e} du Foresto}, {ten Brummelaar}, {McAlister}, {Ridgway},
  {Sturmann}, {Sturmann}, \& {Turner}}]{Merand_2015_12_0}
{M{\'e}rand}, A., {Kervella}, P., {Breitfelder}, J., {et~al.} 2015, \aap, 584,
  A80

\bibitem[{{M{\'e}rand} {et~al.}(2006){M{\'e}rand}, {Kervella}, {Coud{\'e} Du
  Foresto}, {Perrin}, {Ridgway}, {Aufdenberg}, {Ten Brummelaar}, {McAlister},
  {Sturmann}, {Sturmann}, {Turner}, \& {Berger}}]{Merand_2006_07_0}
{M{\'e}rand}, A., {Kervella}, P., {Coud{\'e} Du Foresto}, V., {et~al.} 2006,
  \aap, 453, 155

\bibitem[{{Moffett} \& {Barnes}(1984)}]{Moffett_1984_07_0}
{Moffett}, T.~J. \& {Barnes}, III, T.~G. 1984, \apjs, 55, 389

\bibitem[{{Monnier} {et~al.}(2004){Monnier}, {Berger}, {Millan-Gabet}, \& {ten
  Brummelaar}}]{Monnier_2004_10_0}
{Monnier}, J.~D., {Berger}, J.-P., {Millan-Gabet}, R., \& {ten Brummelaar},
  T.~A. 2004, in SPIE Conference Series, ed. {W.~A.~Traub}, Vol. 5491, 1370

\bibitem[{{Monnier} {et~al.}(2018){Monnier}, {Le Bouquin}, {Anugu}, {Kraus},
  {Setterholm}, {Ennis}, {Lanthermann}, {Jocou}, \& {ten
  Brummelaar}}]{Monnier_2018_07_1}
{Monnier}, J.~D., {Le Bouquin}, J.-B., {Anugu}, N., {et~al.} 2018, in SPIE
  Conference Series, Vol. 10701, 1070122

\bibitem[{{Monnier} {et~al.}(2007){Monnier}, {Zhao}, {Pedretti}, {Thureau},
  {Ireland}, {Muirhead}, {Berger}, {Millan-Gabet}, {Van Belle}, {ten
  Brummelaar}, {McAlister}, {Ridgway}, {Turner}, {Sturmann}, {Sturmann}, \&
  {Berger}}]{Monnier_2007_07_0}
{Monnier}, J.~D., {Zhao}, M., {Pedretti}, E., {et~al.} 2007, Science, 317, 342

\bibitem[{{Monson} \& {Pierce}(2011)}]{Monson_2011_03_0}
{Monson}, A.~J. \& {Pierce}, M.~J. 2011, \apjs, 193, 12

\bibitem[{{Neilson} {et~al.}(2011){Neilson}, {Cantiello}, \&
  {Langer}}]{Neilson_2011_05_0}
{Neilson}, H.~R., {Cantiello}, M., \& {Langer}, N. 2011, \aap, 529, L9

\bibitem[{{Nguyen} {et~al.}(2022){Nguyen}, {Costa}, {Girardi}, {Volpato},
  {Bressan}, {Chen}, {Marigo}, {Fu}, \& {Goudfrooij}}]{Nguyen_2022_09_0}
{Nguyen}, C.~T., {Costa}, G., {Girardi}, L., {et~al.} 2022, \aap, 665, A126

\bibitem[{{Pecaut} \& {Mamajek}(2013)}]{Pecaut_2013_09_0}
{Pecaut}, M.~J. \& {Mamajek}, E.~E. 2013, \apjs, 208, 9

\bibitem[{{Pilecki}(2024)}]{Pilecki_2024_07_0}
{Pilecki}, B. 2024, \apjl, 970, L14

\bibitem[{{Pilecki} {et~al.}(2018){Pilecki}, {Gieren}, {Pietrzy{\'n}ski},
  {Thompson}, {Smolec}, {Graczyk}, {Taormina}, {Udalski}, {Storm}, {Nardetto},
  {Gallenne}, {Kervella}, {Soszy{\'n}ski}, {G{\'o}rski}, {Wielg{\'o}rski},
  {Suchomska}, {Karczmarek}, \& {Zgirski}}]{Pilecki_2018_07_0}
{Pilecki}, B., {Gieren}, W., {Pietrzy{\'n}ski}, G., {et~al.} 2018, \apj, 862,
  43

\bibitem[{{Pilecki} {et~al.}(2021){Pilecki}, {Pietrzy{\'n}ski}, {Anderson},
  {Gieren}, {Taormina}, {Narloch}, {Evans}, \& {Storm}}]{Pilecki_2021_04_0}
{Pilecki}, B., {Pietrzy{\'n}ski}, G., {Anderson}, R.~I., {et~al.} 2021, \apj,
  910, 118

\bibitem[{{Pilecki} {et~al.}(2024){Pilecki}, {Thompson}, {Espinoza-Arancibia},
  {Hajdu}, {Gieren}, {Taormina}, {Pietrzy{\'n}ski}, {Narloch}, {Bono},
  {Gallenne}, {Kervella}, {Wielg{\'o}rski}, {Zgirski}, {Graczyk}, {Karczmarek},
  \& {Evans}}]{Pilecki_2024_06_0}
{Pilecki}, B., {Thompson}, I.~B., {Espinoza-Arancibia}, F., {et~al.} 2024,
  \aap, 686, A263

\bibitem[{{Proffitt} {et~al.}(2017){Proffitt}, {Evans}, {Winston}, {Gallenne},
  \& {Kervella}}]{Proffitt_2017_09_0}
{Proffitt}, C.~R., {Evans}, N.~R., {Winston}, E.~M., {Gallenne}, A., \&
  {Kervella}, P. 2017, in European Physical Journal Web of Conferences, Vol.
  152, 04003

\bibitem[{{Sandage} {et~al.}(2004){Sandage}, {Tammann}, \&
  {Reindl}}]{Sandage_2004_09_0}
{Sandage}, A., {Tammann}, G.~A., \& {Reindl}, B. 2004, \aap, 424, 43

\bibitem[{{Setterholm} {et~al.}(2023){Setterholm}, {Monnier}, {Le Bouquin},
  {Anugu}, {Ennis}, {Jocou}, {Ibrahim}, {Kraus}, {Anderson}, {Chhabra},
  {Codron}, {Farrington}, {Flores}, {Gardner}, {Gutierrez}, {Lanthermann},
  {Majoinen}, {Mortimer}, {Schaefer}, {Scott}, {ten Brummelaar}, \&
  {Vargas}}]{Setterholm_2023_04_0}
{Setterholm}, B.~R., {Monnier}, J.~D., {Le Bouquin}, J.-B., {et~al.} 2023,
  Journal of Astronomical Telescopes, Instruments, and Systems, 9, 025006

\bibitem[{{Storm} {et~al.}(2011){Storm}, {Gieren}, {Fouqu{\'e}}, {Barnes},
  {Pietrzy{\'n}ski}, {Nardetto}, {Weber}, {Granzer}, \&
  {Strassmeier}}]{Storm_2011_10_0}
{Storm}, J., {Gieren}, W., {Fouqu{\'e}}, P., {et~al.} 2011, \aap, 534, A94

\bibitem[{{Szabados}(1977)}]{Szabados_1977_01_0}
{Szabados}, L. 1977, Commun. of the Konkoly Observatory Hungary, 70, 1

\bibitem[{{Szil{\`a}di} {et~al.}(2018){Szil{\`a}di}, {Vink{\`o}}, \&
  {Szabados}}]{Sziladi_2018_06_0}
{Szil{\`a}di}, K., {Vink{\`o}}, J., \& {Szabados}, L. 2018, \actaa, 68, 111

\bibitem[{{ten Brummelaar} {et~al.}(2005){ten Brummelaar}, {McAlister},
  {Ridgway}, {Bagnuolo}, {Turner}, {Sturmann}, {Sturmann}, {Berger}, {Ogden},
  {Cadman}, {Hartkopf}, {Hopper}, \& {Shure}}]{ten-Brummelaar_2005_07_0}
{ten Brummelaar}, T.~A., {McAlister}, H.~A., {Ridgway}, S.~T., {et~al.} 2005,
  \apj, 628, 453

\bibitem[{{Trahin}(2019)}]{Trahin_2019_11_0}
{Trahin}, B. 2019, PhD thesis, Universit\'e PSL
  (https://hal.archives-ouvertes.fr/tel-02372923)

\bibitem[{{Trahin} {et~al.}(2021){Trahin}, {Breuval}, {Kervella}, {M{\'e}rand},
  {Nardetto}, {Gallenne}, {Hocd{\'e}}, \& {Gieren}}]{Trahin_2021_12_5}
{Trahin}, B., {Breuval}, L., {Kervella}, P., {et~al.} 2021, \aap, 656, A102

\bibitem[{{Wahlgren} \& {Evans}(1998)}]{Wahlgren_1998_04_0}
{Wahlgren}, G.~M. \& {Evans}, N.~R. 1998, \aap, 332, L33

\bibitem[{{Wesselink}(1946)}]{Wesselink_1946_01_0}
{Wesselink}, A.~J. 1946, \bain, 10, 91

\bibitem[{{Wilson} {et~al.}(1989){Wilson}, {Carter}, {Barnes}, {van Citters},
  \& {Moffett}}]{Wilson_1989_04_0}
{Wilson}, T.~D., {Carter}, M.~W., {Barnes}, Thomas~G., I., {van Citters},
  G.~Wayne, J., \& {Moffett}, T.~J. 1989, \apjs, 69, 951

\bibitem[{{Wright} \& {Howard}(2009)}]{Wright_2009_05_0}
{Wright}, J.~T. \& {Howard}, A.~W. 2009, \apjs, 182, 205

\bibitem[{{Zucker} \& {Alexander}(2007)}]{Zucker_2007_01_0}
{Zucker}, S. \& {Alexander}, T. 2007, \apjl, 654, L83

\end{thebibliography}
 
 
\begin{appendix}
\onecolumn
\section{Comparison of SU~Cyg Ba spectrum with models}
\label{appendix__Comparison_of_SU_Cyg_B_spectrum_with_models}

\begin{figure*}[!h]
	\centering
	\resizebox{\hsize}{!}{\includegraphics{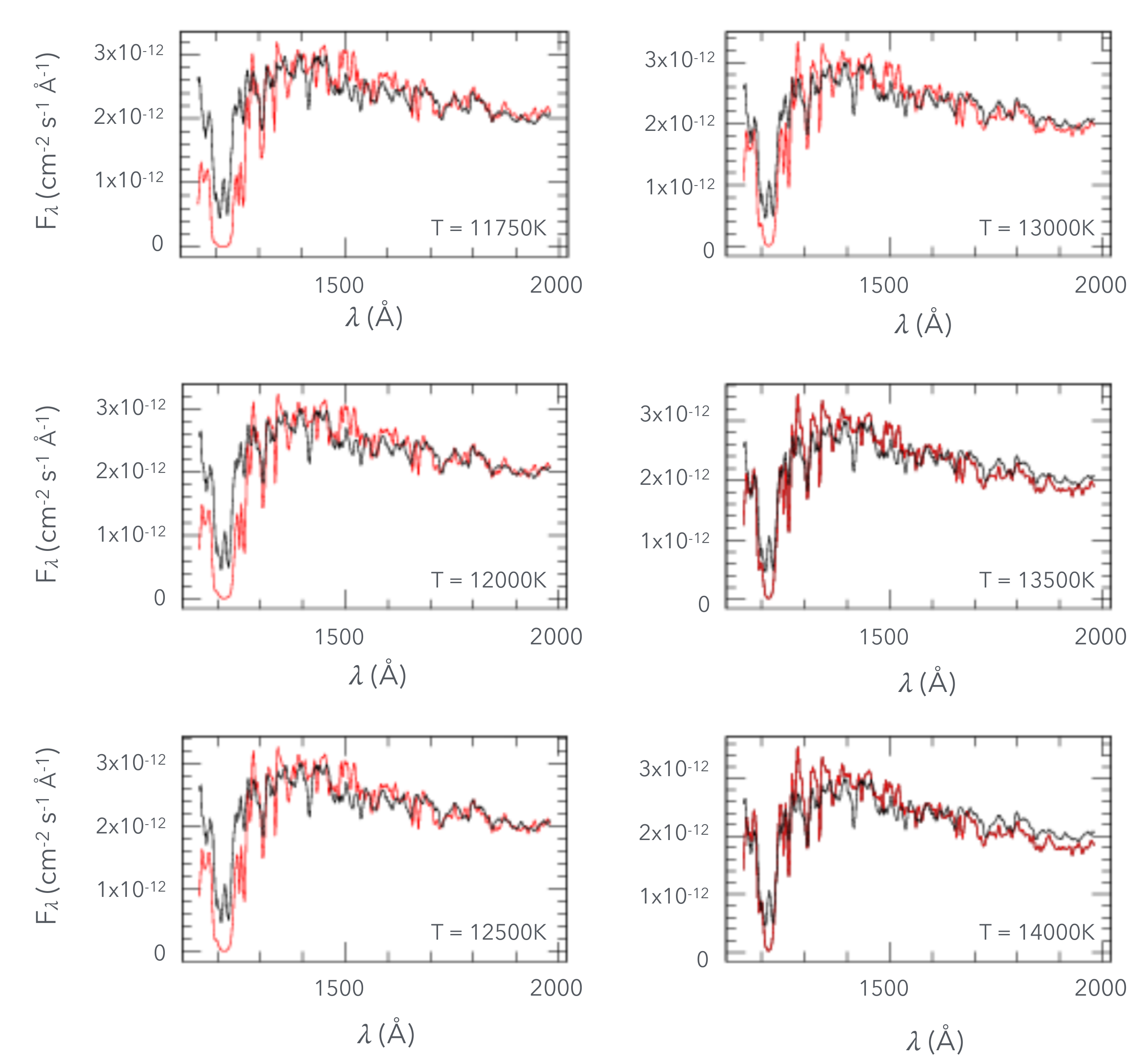}}
	\caption{Comparison between BOSZ models and the STIS  spectrum dereddened for $E(B-V) = 0.109$\,mag. Models are in red and spectrum in black. The temperature for the models is listed within each plot.}
	\label{figure_sm_spec_and_model}
\end{figure*}

\begin{figure*}[!h]
	\centering
	\resizebox{.65\hsize}{!}{\includegraphics{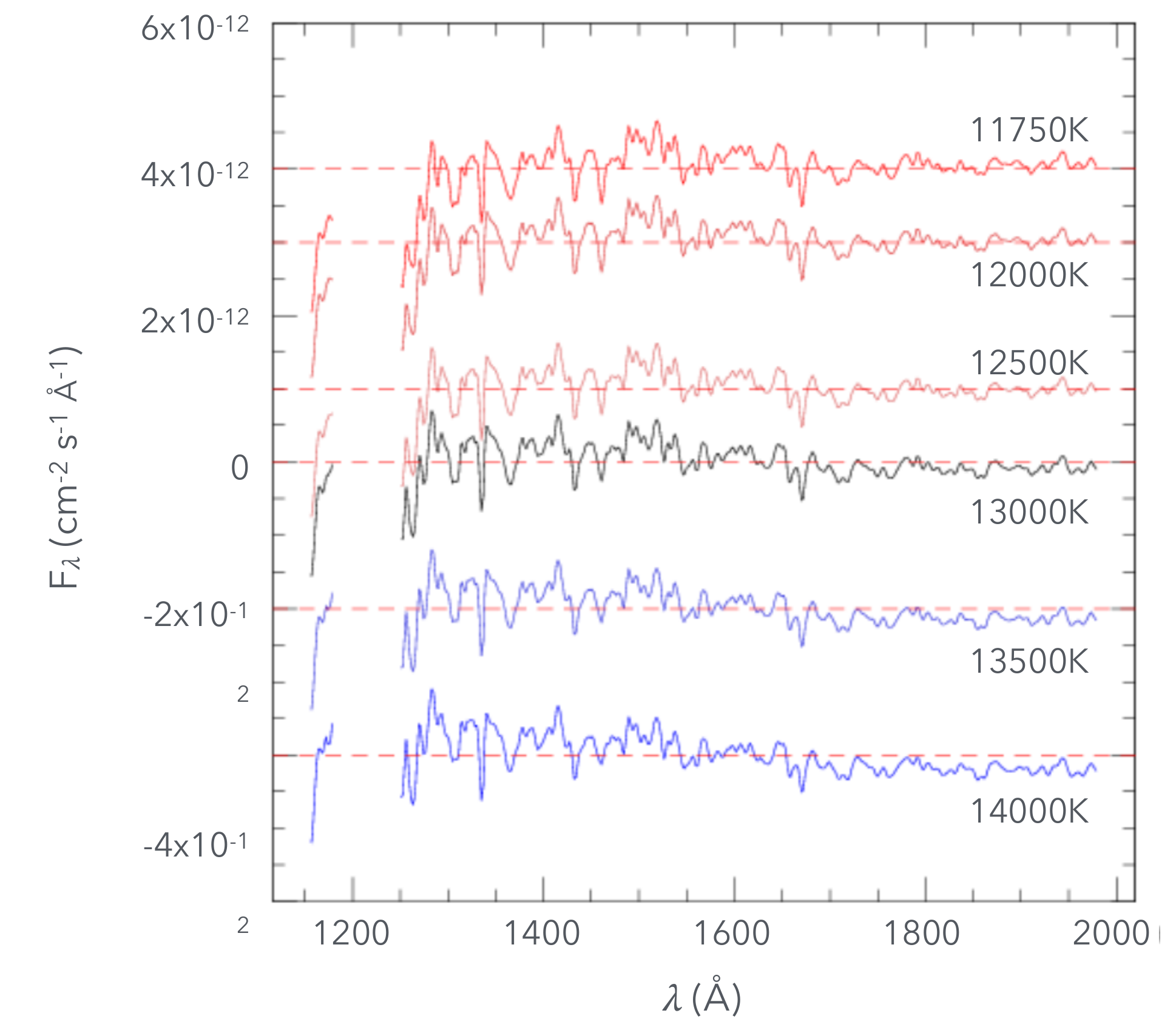}}
	\caption{The difference between the model and the spectrum for $E(B-V) = 0.109$\,mag. The temperatures for the models is written on each plot. The wavelengths between $1180-1250\,\AA$ are omitted because of contamination by interstellar Ly$\alpha$ absorption.}
	\label{figure_sm_diff_single}
\end{figure*}

\begin{figure*}[!h]
	\centering
	\resizebox{.65\hsize}{!}{\includegraphics{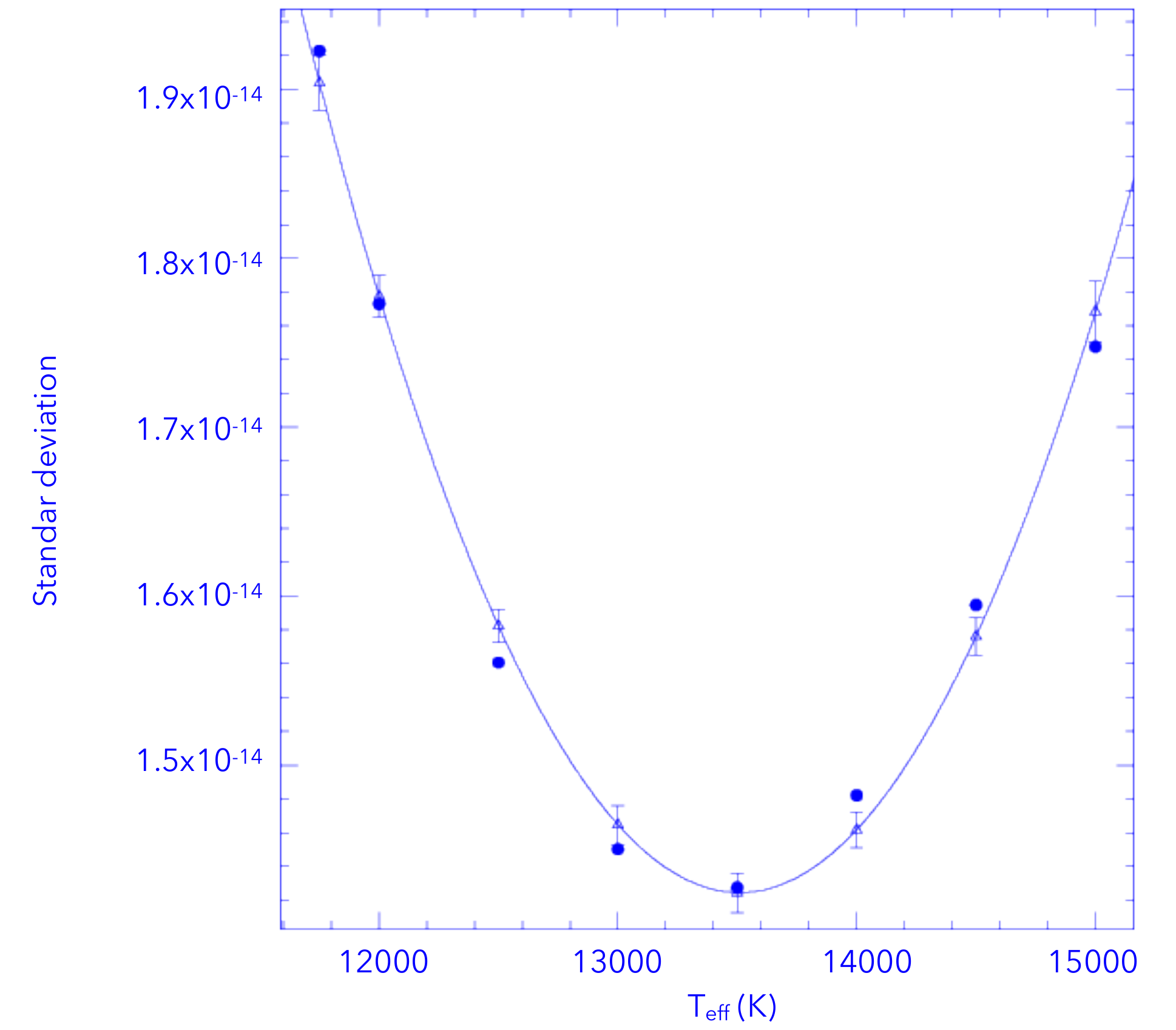}}
	\caption{The standard deviations from the spectrum-model comparison as the temperature of the models is changed. Dots: the standard deviation; triangles: the parabola fit.}
	\label{figure_sm}
\end{figure*}

The spectrum of SU~Cyg~Ba is provided in IUE spectrum SWP 14773. In the wavelength region $1150-1900\,\AA$ the Cepheid SU~Cyg~A contributes negligible flux, as does the less massive cooler companion SU~Cyg~Bb. The spectrum is dereddened by $E(B-V) = 0.109$\,mag, and also corrected to the HST STIS flux scale. The spectrum is compared with BOSZ atmospheres in a series of temperatures, as shown in Fig.~\ref{figure_sm_spec_and_model}. The differences between the spectrum and the models is shown Fig.~\ref{figure_sm_diff_single}. The standard deviations from Fig.~\ref{figure_sm_diff_single} are shown in Fig.~\ref{figure_sm} as a function of the model temperature. The temperature found from the parabola in Fig.~\ref{figure_sm} is  $13500\pm 920$\,K

From the absolute magnitude difference between SU~Cyg~A and B and the colour of Ba, the contribution to the combined magnitude from the companion is found to be 0.08\,mag.  The $E(B-V)$ is also recomputed from corrected colours using the appropriate formula from \citet{Fernie_1990_01_0}. The small change in $E(B-V)$ resulted in a negligible change in the temperature of the companion Ba.  

\end{appendix}

\end{document}